%% file: svjourn3-epjc.tex
\documentclass[pdftex,twocolumn,epjc3]{svjour3}          

\RequirePackage[T1]{fontenc}

\smartqed  

\RequirePackage{graphicx}
\RequirePackage{mathptmx}      
\RequirePackage{flushend}
\RequirePackage[numbers,sort&compress]{natbib}
\RequirePackage[colorlinks,citecolor=blue,urlcolor=blue,linkcolor=blue]{hyperref}
\usepackage{amsmath,amssymb}
\usepackage{xspace}
\usepackage{lipsum}
\usepackage{upgreek}
\usepackage{float}
\usepackage{bbold}
\usepackage{placeins}
\usepackage{lineno}
\usepackage{microtype} 

\journalname{Eur. Phys. J.}

\begin{document}
\input{commands.tex}
\title{Novel model for particle emission in small collision systems}

\author{Dimitar Mihaylov\thanksref{e1,addr1}
        \and
        Jaime Gonz\'alez Gonz\'alez\thanksref{addr1}
}

\thankstext{e1}{e-mail: dimitar.mihaylov@mytum.de}

\institute{Physics Department, Technische Universit\"at M\"unchen, James-Franck-Str., 85748 Garching, Germany\label{addr1}
}

\date{Received: date / Accepted: date}

\maketitle

\begin{abstract}

Collider experiments provide an opportunity to produce particles at close distances and momenta. The measured correlation functions between particles can provide information on both the effective emission source and the interaction potential. In recent years, experiments at the LHC have shown that precision studies of the strong interaction are possible using correlation techniques, provided a good handle on the source function. The current work presents a new numerical framework called \textit{\textbf{C}ommon \textbf{E}mission in \textbf{CA}TS} (CECA), capable of simulating the effective emission source of an N-body system based on the properties of the single particles. The framework differentiates between primordial particle emission and particle production through resonances, allowing to verify the hypothesis proposed by the ALICE collaboration that a common baryon--baryon emission source is present in small collision systems. The new framework is used to analyze ALICE data on \pP and \pL correlations and compare the results to previous studies based on the common emission source scenario. It is demonstrated that the best fit to the \pL correlation data is obtained using a scattering length of $1.15\pm0.07$~fm in the S=1 channel.
\\
The new CECA framework provides an essential tool for precision studies in two-body systems and a consistent description of the source function in many-body systems. 
\end{abstract}

\input{Chapters/Introduction.tex}

\input{Chapters/Femtoscopy.tex}

\input{Chapters/CECA.tex}

\input{Chapters/Analysis.tex}
\input{Chapters/Summary.tex}

\begin{acknowledgements}
We would like to thank Dr. Valentina Mantovani Sarti and Prof. Laura Fabbietti for their constant support and many discussions during the analysis of the \pP and \pL correlations. Further, the discussions with Dr. Johann Haidenbauer were essential in understanding the details of the \pL interaction, which enabled the physics interpretation of the results. Finally, we would like to express our gratitude to the ALICE collaboration, as our experience gained as members allowed us to make better use of the existing data.
\end{acknowledgements}

\vspace{-0.45cm}

\appendix
\input{Chapters/Appendix.tex}

\newpage
~
\newpage
~
\newpage
~
\newpage
~
\newpage

\label{Bibliography}
\bibliographystyle{unsrtnat}
\bibliography{bibliography.bib}

\end{document}

%% file: commands.tex
%

\newcommand{\pp}           {pp\xspace}
\newcommand{\ppbar}        {\mbox{$\mathrm {p\overline{p}}$}\xspace}
\newcommand{\XeXe}         {\mbox{Xe--Xe}\xspace}
\newcommand{\PbPb}         {\mbox{Pb--Pb}\xspace}
\newcommand{\pPb}          {\mbox{p--Pb}\xspace}
\newcommand{\AuAu}         {\mbox{Au--Au}\xspace}
\newcommand{\dAu}          {\mbox{d--Au}\xspace}

\newcommand{\snn}          {\ensuremath{\sqrt{s_{\mathrm{NN}}}}\xspace}
\newcommand{\pt}           {\ensuremath{p_{\rm T}}\xspace}
\newcommand{\meanpt}       {$\langle p_{\mathrm{T}}\rangle$\xspace}
\newcommand{\ycms}         {\ensuremath{y_{\rm CMS}}\xspace}
\newcommand{\ylab}         {\ensuremath{y_{\rm lab}}\xspace}
\newcommand{\etarange}[1]  {\mbox{$\left | \eta \right |~<~#1$}}
\newcommand{\yrange}[1]    {\mbox{$\left | y \right |~<~#1$}}
\newcommand{\dndy}         {\ensuremath{\mathrm{d}N_\mathrm{ch}/\mathrm{d}y}\xspace}
\newcommand{\dndeta}       {\ensuremath{\mathrm{d}N_\mathrm{ch}/\mathrm{d}\eta}\xspace}
\newcommand{\avdndeta}     {\ensuremath{\langle\dndeta\rangle}\xspace}
\newcommand{\dNdy}         {\ensuremath{\mathrm{d}N_\mathrm{ch}/\mathrm{d}y}\xspace}
\newcommand{\Npart}        {\ensuremath{N_\mathrm{part}}\xspace}
\newcommand{\Ncoll}        {\ensuremath{N_\mathrm{coll}}\xspace}
\newcommand{\dEdx}         {\ensuremath{\textrm{d}E/\textrm{d}x}\xspace}
\newcommand{\RpPb}         {\ensuremath{R_{\rm pPb}}\xspace}
\newcommand{\mt}           {\ensuremath{m_{\rm T}}\xspace}
\newcommand{\kt}           {\ensuremath{k_{\rm T}}\xspace}
\newcommand{\rd}           {\ensuremath{r_d}\xspace}
\newcommand{\rdv}           {\ensuremath{\vec{r_d}}\xspace}
\newcommand{\h}           {\ensuremath{h}\xspace}
\newcommand{\hv}           {\ensuremath{\vec{h}}\xspace}
\newcommand{\hT}           {\ensuremath{h_\mathrm{T}}\xspace}
\newcommand{\hz}           {\ensuremath{h_\mathrm{z}}\xspace}
\newcommand{\rs}           {\ensuremath{r^*}\xspace}
\newcommand{\rsv}           {\ensuremath{\vec{r}^*}\xspace}
\newcommand{\rc}           {\ensuremath{r_\mathrm{core}}\xspace}
\newcommand{\rcv}           {\ensuremath{\vec{r}_\mathrm{core}}\xspace}
\newcommand{\rcs}           {\ensuremath{r^*_\mathrm{core}}\xspace}
\newcommand{\rcsv}           {\ensuremath{\vec{r}^*_\mathrm{core}}\xspace}
\newcommand{\reff}           {\ensuremath{r_\mathrm{eff}}\xspace}
\newcommand{\ks}           {\ensuremath{k^*}\xspace}
\newcommand{\kst}           {\ensuremath{k^*_\mathrm{true}}\xspace}
\newcommand{\kstar}           {\ensuremath{k^*}\xspace}
\newcommand{\ksv}           {\ensuremath{\vec{k}^*}\xspace}
\newcommand{\Sr}           {\ensuremath{S(r)}\xspace}
\newcommand{\Ck}           {\ensuremath{C(k)}\xspace}
\newcommand{\Srs}           {\ensuremath{S(\rs)}\xspace}
\newcommand{\Srsv}           {\ensuremath{S(\rsv)}\xspace}
\newcommand{\Cks}           {\ensuremath{C(\ks)}\xspace}
\newcommand{\SEks}           {\ensuremath{N(\ks)}\xspace}
\newcommand{\MEks}           {\ensuremath{M(\ks)}\xspace}
\newcommand{\wf}           {\ensuremath{\psi(\vec{\rs},\vec{\ks})}\xspace}
\newcommand{\NN}           {\ensuremath{\mbox{NN}}\xspace}
\newcommand{\pP}           {\ensuremath{\mbox{pp}}\xspace}
\newcommand{\ApAP}         {\ensuremath{\mbox{\pbar\pbar}}\xspace}
\newcommand{\pL}           {\ensuremath{\rm \mbox{p}\Lambda}\xspace}
\newcommand{\Lam} {\ensuremath{\rm\Lambda}\xspace}
\newcommand{\Sig} {\ensuremath{\rm\Sigma}\xspace}
\newcommand{\NS} {\ensuremath{\rm N\Sig}\xspace}
\newcommand{\NL} {\ensuremath{\rm N\Lam}\xspace}
\newcommand{\decpLpL}           {\ensuremath{\mathrm{p}\Lambda}\xspace}
\newcommand{\decpLpSo}           {\ensuremath{\mathrm{p}(\Sigma^0)}\xspace}
\newcommand{\decpLpXi}           {\ensuremath{\mathrm{p}(\Xi)}\xspace}
\newcommand{\decpLpXio}           {\ensuremath{\mathrm{p}(\Xi^0)}\xspace}
\newcommand{\decpLpXim}           {\ensuremath{\mathrm{p}(\Xi^-)}\xspace}
\newcommand{\decpLflat}           {\ensuremath{\mathrm{ff}}\xspace}
\newcommand{\decpfL}           {\ensuremath{\mathrm{p}\tilde{\Lambda}}\xspace}
\newcommand{\decfpL}           {\ensuremath{\tilde{\mathrm{p}}\Lambda}\xspace}
\newcommand{\KpL}           {\ensuremath{K^{+}\Lambda}\xspace}
\newcommand{\pK}           {\ensuremath{\mbox{p}\mathrm{K}}\xspace}
\newcommand{\pKminus}           {\ensuremath{\mbox{p}\mathrm{K}^-}\xspace}
\newcommand{\pKplus}           {\ensuremath{\mbox{p}\mathrm{K}^-}\xspace}
\newcommand{\nAKo}           {\ensuremath{\mbox{n--}\mathrm{\overline{K}}^0}\xspace}
\newcommand{\ApAL}         {\ensuremath{\mbox{\pbar\almb}}\xspace}
\newcommand{\LL}           {\ensuremath{\Lambda\Lambda}\xspace}
\newcommand{\pSo}          {\ensuremath{\rm \mbox{p}\Sigma^{0}}\xspace}
\newcommand{\pXi}          {\ensuremath{\rm \mbox{p}\Xi}\xspace}
\newcommand{\pXio}          {\ensuremath{\rm \mbox{p}\Xi^{0}}\xspace}
\newcommand{\pXim}          {\ensuremath{\rm \mbox{p}\Xi^{-}}\xspace}
\newcommand{\pOmega}          {\ensuremath{\rm\mbox{p\Om}}\xspace}
\newcommand{\pPhi}          {\ensuremath{\mbox{p}\upphi}\xspace}
\newcommand{\pipi}      {\ensuremath{\uppi\uppi}\xspace}
\newcommand{\ppip}           {\ensuremath{\mbox{p}\pip}\xspace}
\newcommand{\ppim}           {\ensuremath{\mbox{p}\pim}\xspace}
\newcommand{\kaka}          {\ensuremath{\mbox{KK}}\xspace}

\newcommand{\pLNS}          {\ensuremath{\mathrm{p}\Lambda\mbox{--}\mathrm{N}\Sigma\xspace}}
\newcommand{\pKmnaKo}          {\ensuremath{\mathrm{pK^{-}}\mbox{--}\mathrm{n\overline{K^{0}}}}\xspace}
\newcommand{\NKcoupling}          {\ensuremath{\mathrm{N\overline{K}}\mbox{--}\mathrm{\uppi\Sigma}\mbox{--}\mathrm{\uppi\Lambda}\xspace}}
\newcommand{\pLtopL}          {\ensuremath{\mathrm{p}\Lambda\mbox{--}\mathrm{p}\Lambda\xspace}}
\newcommand{\NStopL}          {\ensuremath{\mathrm{N}\Sigma\mbox{--}\mathrm{p}\Lambda\xspace}}
\newcommand{\pSotopL}          {\ensuremath{\mathrm{p}\Sigma^0\mbox{--}\mathrm{p}\Lambda\xspace}}
\newcommand{\nSptopL}          {\ensuremath{\mathrm{n}\Sigma^+\mbox{--}\mathrm{p}\Lambda\xspace}}

\newcommand{\nineH}        {$\sqrt{s}~=~0.9$~Te\kern-.1emV\xspace}
\newcommand{\seven}        {$\sqrt{s}~=~7$~Te\kern-.1emV\xspace}
\newcommand{\onethree}        {$\sqrt{s}~=~13$~Te\kern-.1emV\xspace}
\newcommand{\twoH}         {$\sqrt{s}~=~0.2$~Te\kern-.1emV\xspace}
\newcommand{\twosevensix}  {$\sqrt{s}~=~2.76$~Te\kern-.1emV\xspace}
\newcommand{\five}         {$\sqrt{s}~=~5.02$~Te\kern-.1emV\xspace}
\newcommand{\twosevensixnn}{$\sqrt{s_{\mathrm{NN}}}~=~2.76$~Te\kern-.1emV\xspace}
\newcommand{\fivenn}       {$\sqrt{s_{\mathrm{NN}}}~=~5.02$~Te\kern-.1emV\xspace}
\newcommand{\LT}           {L{\'e}vy-Tsallis\xspace}
\newcommand{\lumi}         {\ensuremath{\mathcal{L}}\xspace}
\newcommand{\MeV}  {\ensuremath{\text{Me\kern-.1emV}}\xspace}
\newcommand{\MeVc}  {\ensuremath{\text{Me\kern-.1emV/}c}\xspace}
\newcommand{\MeVcc}  {\ensuremath{\text{Me\kern-.2emV/}c^2}\xspace}
\newcommand{\GeV}  {\ensuremath{\text{Ge\kern-.1emV}}\xspace}
\newcommand{\GeVc}  {\ensuremath{\text{Ge\kern-.1emV/}c}\xspace}
\newcommand{\GeVcc}  {\ensuremath{\text{Ge\kern-.2emV/}c^2}\xspace}
\newcommand{\TeV}  {\ensuremath{\text{Te\kern-.1emV}}\xspace}
\newcommand{\fmc}  {\ensuremath{\text{fm/}c}\xspace}

\newcommand{\ITS}          {\rm{ITS}\xspace}
\newcommand{\TOF}          {\rm{TOF}\xspace}
\newcommand{\ZDC}          {\rm{ZDC}\xspace}
\newcommand{\ZDCs}         {\rm{ZDCs}\xspace}
\newcommand{\ZNA}          {\rm{ZNA}\xspace}
\newcommand{\ZNC}          {\rm{ZNC}\xspace}
\newcommand{\SPD}          {\rm{SPD}\xspace}
\newcommand{\SDD}          {\rm{SDD}\xspace}
\newcommand{\SSD}          {\rm{SSD}\xspace}
\newcommand{\TPC}          {\rm{TPC}\xspace}
\newcommand{\TRD}          {\rm{TRD}\xspace}
\newcommand{\VZERO}        {\rm{V0}\xspace}
\newcommand{\VZEROA}       {\rm{V0A}\xspace}
\newcommand{\VZEROC}       {\rm{V0C}\xspace}
\newcommand{\Vdecay} 	   {\ensuremath{V^{0}}\xspace}

\newcommand{\ee}           {\ensuremath{e^{+}e^{-}}} 
\newcommand{\pip}          {\ensuremath{\uppi^{+}}\xspace}
\newcommand{\pim}          {\ensuremath{\uppi^{-}}\xspace}
\newcommand{\kp}          {\ensuremath{\rm{K}^{+}}\xspace}
\newcommand{\km}          {\ensuremath{\rm{K}^{-}}\xspace}
\newcommand{\proton}       {\ensuremath{\rm{p}}\xspace}
\newcommand{\pbar}         {\ensuremath{\rm\overline{p}}\xspace}
\newcommand{\kzero}        {\ensuremath{{\rm K}^{0}_{\rm{S}}}\xspace}
\newcommand{\lmb}          {\ensuremath{\Lambda}\xspace}
\newcommand{\almb}         {\ensuremath{\overline{\Lambda}}\xspace}
\newcommand{\So}           {\ensuremath{\Sigma^{0}}\xspace}
\newcommand{\Om}           {\ensuremath{\Omega^-}\xspace}
\newcommand{\Mo}           {\ensuremath{\overline{\Omega}^+}\xspace}
\newcommand{\X}            {\ensuremath{\Xi^-}\xspace}
\newcommand{\Ix}           {\ensuremath{\overline{\Xi}^+}\xspace}
\newcommand{\Xis}          {\ensuremath{\Xi^{\pm}}\xspace}
\newcommand{\Oms}          {\ensuremath{\Omega^{\pm}}\xspace}

\newcommand{\Cth}           {C_\mathrm{th}}
\newcommand{\Cexp}           {C_\mathrm{exp}}
\newcommand{\CF}           {C(\ks)}

\newcommand{\pprot}            {\ensuremath{\proton\mbox{--}\proton}\xspace}

\newcommand{\LP}{\ensuremath{\rm p \Lambda}\xspace}
\newcommand{\PL}{\ensuremath{\rm p \Lambda}\xspace}
\newcommand{\PXim}{\ensuremath{\rm p \Xi^-}\xspace}
\newcommand{\PXiZ}{\ensuremath{\rm p \Xi^0}\xspace}
\newcommand{\PSigZ}{\ensuremath{\rm p\Sigma^0}\xspace}

\newcommand{\SN}{\ensuremath{\rm N \Sigma}\xspace}
\newcommand{\LN}{\ensuremath{\rm N \Lambda}\xspace}
\newcommand{\SNLN}{\ensuremath{\rm N \Sigma\leftrightarrow N \Lambda}\xspace}
\newcommand{\SNtoLN}{\ensuremath{\rm N \Sigma\rightarrow N \Lambda}\xspace}
\newcommand{\Kmp}{\ensuremath{\rm K^- p}\xspace}

\newcommand{\diff}           {\mathrm{d}}
\newcommand{\cosine}           {\mathrm{cos}}

\newcommand{\Chieft}           {\ensuremath{\rm \chi EFT}\xspace}
\newcommand{\nsig}           {\ensuremath{n\sigma}\xspace}

%% file: Chapters/Introduction.tex
\section{Introduction}\label{sec:Intro}
The use of correlation techniques in particle physics dates to the 1960s, tracing its origin in the \textit{Hanbury Brown and Twiss effect} (HBT), developed in astronomy~\cite{HanburyBrown:1954amm}. The essence of the HBT method is to relate the wave interference with the spatial distribution of the signal emitter, by measuring and dividing the incoming intensities of a correlated and an uncorrelated signal. In particle physics, the HBT effect has commonly been applied to two-body correlations of particles produced in collider experiments. The methodology is referred to as \textit{femtoscopy}, as the correlation signal is generated from the underlying hadronization (emission) process and the subsequent \textit{final state interaction} (FSI), both of which happen on the femtometer scale~\cite{Lisa:2005dd}. In particular, particle pairs with low relative momentum \ks, where the asterisk denotes the pair rest frame, are significantly influenced by quantum effects, generating a strong correlation signal. This feature has been extensively used to test the so-called \textit{particle emission source}, which is an effective parameterization of the spatial components of the particles at the time of their hadronization. Any subsequent modification of the momentum is associated with the FSI, related to the strong and Coulomb forces, as well as the symmetrization properties of the wave function. Consequently, two-particle correlation functions measured at low \ks contain information both related to the emission source and to the low energy scattering~\cite{Lisa:2005dd,STAR:2004qya,ALICE:Run1,ALICE:pK_prl,ALICE:pOmega}. 

\noindent In recent years, the ALICE collaboration demonstrated that small collision systems, \pp in particular, result in a common emission source for all baryon--baryon pairs~\cite{ALICE:Run1}. This source is obtained using \pP correlations as a benchmark, since for these pairs the interaction is well known~\cite{Wiringa:AV18}. In particular, a detailed analysis of the contributions of short-lived resonances to the source properties has been carried out~\cite{ALICE:Source,Wiedemann:1996ig}. Additionally, it has been shown that in small collision systems the source size strongly depends on the \textit{transverse mass} (\mt) of the pair, as previously observed in \textit{heavy-ion} (HI) collisions. This behavior is commonly associated with collective effects, such as radial flow~\cite{Wiedemann:1999qn,Heinz:1999rw}, but the \mt scaling in \pp collisions is not yet fully understood or modeled. This issue will be addressed in the present work. 

\noindent The main motivation for studying the emission source in small collision systems is the possibility to measure particle pairs of poorly constrained interaction and to access the low energy scattering properties through the correlation function~\cite{ALICE:Run1,ALICE:pK_prl,ALICE:pOmega}. Most of the recent analyses of two-particle correlations in the context of the FSI were performed with the \textit{"Correlation Analysis tool using the Schr\"odinger Equation"} (CATS)~\cite{Mihaylov:2018rva}. The latter is a numerical framework capable of evaluating the correlation function by taking as input either an interaction potential or a two-particle wave function, as well as an emission source of any form.\\ The current work presents a numerical extension to the CATS framework, aimed at an improved modeling of the emission source. The \textit{"Common Emission in CATS"} (CECA) framework relies on simulating single-particle emission including spatial-momentum correlations to generate the \mt scaling. The model has been validated by re-analyzing the ALICE \pP and \pL correlation functions~\cite{ALICE:Source}, measured differentially in bins of \mt, and by employing the assumption of a common emission source for both species. Further, it is demonstrated that a reduced attraction within the \pL interaction leads to an overall better description of the data. This finding is consistent with the \mt integrated analysis of the \pL system~\cite{ALICE:pL}, in which a $\approx 3\sigma$ tension between the state-of-the-art \textit{chiral effective field theory} (\Chieft)~\cite{Haidenbauer:NLO19} and the data is observed. It is likely that the insufficient statistical significance of the scattering data used to constrain the \Chieft is responsible for this deviation. The \pL interaction is an important ingredient for the construction of a realistic nuclear equation of state for dense nuclear matter, which would help to study the composition of neutron stars~\cite{Kaiser:2004fe,Lonardoni:2014bwa,Gerstung:2020ktv}. The present work pins down the properties of the \pL interaction by using the ALICE \mt differential data and by applying the CATS framework, assuming a common source function for protons and \Lam baryons modeled by CECA.

\noindent This paper is organized as follows. In Section~\ref{sec:Femto} the basics of femtoscopy are explained, where sub-section~\ref{sec:Femto:Source} provides detailed information on the properties of the emission source relevant to the presented work.

\noindent Section~\ref{sec:CECA} explains the working principles of the CECA frameworks, providing information on the individual parameters and several examples to illustrate their effect on the emission profile and the associated \mt dependence.

\noindent In Section~\ref{sec:pp_pLCorr} the CECA framework is applied to ALICE data, in order to find a parameterization of the source capable of modeling both the \pP and \pL correlations for all differential measurements in \mt. A common emission source for both the protons and \Lam particles is assumed. 
Further, the properties of the \pL interaction will be examined, and it will be demonstrated that a weaker two-body attraction, compared to currently accepted values, is required in order to achieve the best fit to the data.

\noindent In Section~\ref{sec:Summary} a summary and outlook are presented, commenting on the future applications of the framework.

%% file: Chapters/Femtoscopy.tex
\section{Femtoscopy}\label{sec:Femto}
\subsection{Overview}\label{sec:Femto:Overview}
In two-body systems, the femtoscopic formalism is based on the Koonin-Pratt relation~\cite{Lisa:2005dd}
\begin{equation}\label{eq:KooninPratt_Simple}
C(\ks)=\int S(\ksv,\rsv)\left|\Psi(\ksv,\rsv)\right|^2d^3\rs,
\end{equation}
where \rs is the relative distance between the particles at the time of their effective emission, $S(\ksv,\rsv)$ is the source function and $\Psi(\ksv,\rsv)$ is the wave function of the relative motion of the particle pair. The source function is an effective parameterization of the properties of the particle emission. Typically, the \ks dependence is ignored, while only the radial part of the spatial components is considered. This is also assumed in the present work. In the specific case of a Gaussian profile, the source function becomes 
\begin{equation}\label{eq:GaussSource}
 S(\rs) = \frac{1}{(4\pi\reff^2)^{3/2}}\exp{\left(-\frac{r^{*2}}{4\reff^2}\right)},
\end{equation}
where \reff is the effective size of the Gaussian source. Eq.~\ref{eq:GaussSource} is valid under the assumption of independent particle emission, where each of the $x$-, $y$- and $z$-components of the single particles are described by a normal distribution with a standard deviation of \reff. Typical values in HI collisions range between 4 and 10~fm, depending on the centrality, while in \pp collisions the source is significantly smaller ($\reff\sim1$~fm).

\noindent From a statistical point of view, the correlation function is the ratio of momenta distributions of correlated pairs $N(\ks)$ and the factorized uncorrelated probability $M(\ks)$. Experimentally, $N(\ks)$ can be obtained by building the pairs from particles stemming from the same collisions (events), while $M(\ks)$ is constructed from uncorrelated particles produced in different (mixed) events. This definition is equivalent to Eq.~\ref{eq:KooninPratt_Simple} if a matching normalization condition is used. In practice, the source function is treated as a \textit{probability density function} (PDF) and normalized to unity over the full spatial integration. In the absence of an interaction, the wave function corresponds to a free wave, which leads to $C(\ks)=1$. In the presence of FSI the correlation function will converge asymptotically towards unity, but it will be enhanced at small \ks in the case of an attractive interaction and depleted in the case of a repulsion. Consequently, $C(\ks)=N(\ks)/M(\ks)\approx 1$ at large $\ks$, which is the typical normalization condition employed in experimental measurements. Nevertheless, due to the likely presence of non-FSI correlations, it is possible that data normalization is systematically biased.

The Koonin-Pratt relation (Eq.~\ref{eq:KooninPratt_Simple}) is valid under the assumption that all particles stem from the emission source and are subject to FSI. In an experimental environment, there are multiple sources of particle production, such as decay products from long-lived ($c\tau\gg$fm) resonances, thus the measurement contains residual correlations~\cite{ALICE:Run1}. These correlations are a contamination to the genuine correlation signal. The total correlation function becomes
\begin{equation}\label{eq:LamPar}
C_\mathrm{tot}(\ks)=\sum_i \lambda_i C_i(\ks),
\end{equation}
where the \textit{$\lambda_i$ parameters} represent the weights of the different contributions. By convention, \textit{i}=0 corresponds to the genuine correlation. In some cases, the genuine signal can further be factorized into several contributions, such as in the presence of multiple spin channels. For example, the \pL system can be either in spin singlet (S=0) or spin triplet (S=1) state, and the genuine correlation function is 
\begin{equation}\label{eq:Spin}
C_\mathrm{gen}(\ks)=\frac{1}{4} C_\mathrm{S=0}(\ks) + \frac{3}{4} C_\mathrm{S=1}(\ks),
\end{equation}
where the weights are derived from the corresponding spin degeneracy~\cite{Mihaylov:2018rva}.

\subsection{Emission source}\label{sec:Femto:Source}

The physics interpretation of the emission source is the effective point in space-time when the FSI between the formed hadrons sets in. In HI collisions this is often associated with the thermal freeze-out and, as such, the emission source and its properties depend on the investigated species due to the re-scattering occurring during the system\textquotesingle s expansion ($\approx 10$ fm$/$c). By contrast, in small collision systems, such as \pp, the hadron production is predominantly linked to hard scattering and parton fragmentation processes, occurring at timescales below 1 fm$/$c. 
These premises constitute the foundations on which modeling of the emitting source, common to all particle species, can be achieved to allow the investigation of the interaction through the measurements of the correlation function.
The ALICE collaboration presented a data-driven approach, called the \textit{Resonance Source Model} (RSM), to describe the emitting source in \pp collisions anchored to the measurement of \pP pairs, in which the interaction is already well constrained from scattering and nuclear data~\cite{ALICE:Source}.
The assumption of the RSM is the existence of an emitting source common for all particles, composed of a Gaussian core of size $r_{\mathrm{core}}$, from which primordial particles are emitted, and an exponential tail coming from the strong decays of resonances, with lifetime $c\tau \lesssim 10$~fm, into the pair of interest. The yields and the kinematics of the resonances are estimated using thermal model predictions~\cite{AndronicSHM} and the EPOS transport model~\cite{EPOS:2013ria}, respectively.
The core radius is the only parameter of the model and it is extracted from a fit to the \pP correlations measured for different \mt ranges of the pairs. The transverse mass is defined as
\begin{equation}\label{eq:mT_correct}
    \mt = \frac{1}{2}\cdot\sqrt{M_\mathrm{inv}^2+(p_\mathrm{T,1}+p_\mathrm{T,2})^2},
\end{equation}
where $M_\mathrm{inv}$ is the invariant mass of the pair, and $p_\mathrm{T,1}$ ($p_\mathrm{T,2}$) are the transverse momenta in the laboratory frame of reference of the first (second) particle. For particle pairs of small relative momenta and similar masses Eq.~\ref{eq:mT_correct} can be transformed into the approximate relation 
\begin{equation}\label{eq:mT_approx}
    \mt \approx \sqrt{m_\mathrm{avg}^2+k_\mathrm{T}^2},
\end{equation}
where $k_\mathrm{T}$ is the average transverse momentum of the two particles and $m_\mathrm{avg}$ is their average mass. The obtained $r_{\mathrm{core}}$ shows a scaling trend as a function of \mt, typically observed in HI experiments and related to the presence of space-momentum correlations in the initial stages of the collisions~\cite{mTscal1,mTscal2,mTscal3,mTscal4}. Further, the ALICE collaboration demonstrates that with the proper inclusion of the strongly decaying resonances, the obtained $r_{\mathrm{core}}$ in \pL measurements is identical to the \pP results, supporting the existence of a common emitting source for baryons in \pp collisions. 
The work in~\cite{ALICE:Source}, and the possibility to anchor the modeling of the emitting source to data using the RSM, allowed to study several interactions in the strange and charm sectors with correlation measurements in \pP collisions at the LHC~\cite{Fabbietti:2020bfg,ALICE:pL,ALICE:BBar,ALICE:LXi,ALICE:pD,ALICE:pKCoupled,ALICE:pOmega,ALICE:pphi}.\\
The full dynamics of the particle propagation from the collision point to the effective point of emission was not present in the RSM approach. This resulted in a failure to reproduce certain observables, in particular the \mt scaling of the source size.
Additionally, the RSM approach does not include Lorentz boost effects, which might introduce further deformations to the measured source radii.\\
These shortcomings of the RSM are addressed within the CECA framework and details are provided in the next section.

%% file: Chapters/CECA.tex
\section{Common Emission in CATS (CECA)}\label{sec:CECA}

The hypothesis of a common emission source in small collision systems is pivotal for performing precision studies on the strong interaction using correlations. Despite the convincing evidence of a common emission in \pp collisions, provided by the ALICE collaboration~\cite{ALICE:Source}, there is a general lack of understanding of its properties. In particular, such a scenario has so far only been tested for baryon pairs, and the observed \mt scaling is not currently reproduced by transport models~\cite{Horst:2023oti}. 

\begin{figure}[h!]
    \centering
    \includegraphics[width=0.45\textwidth]{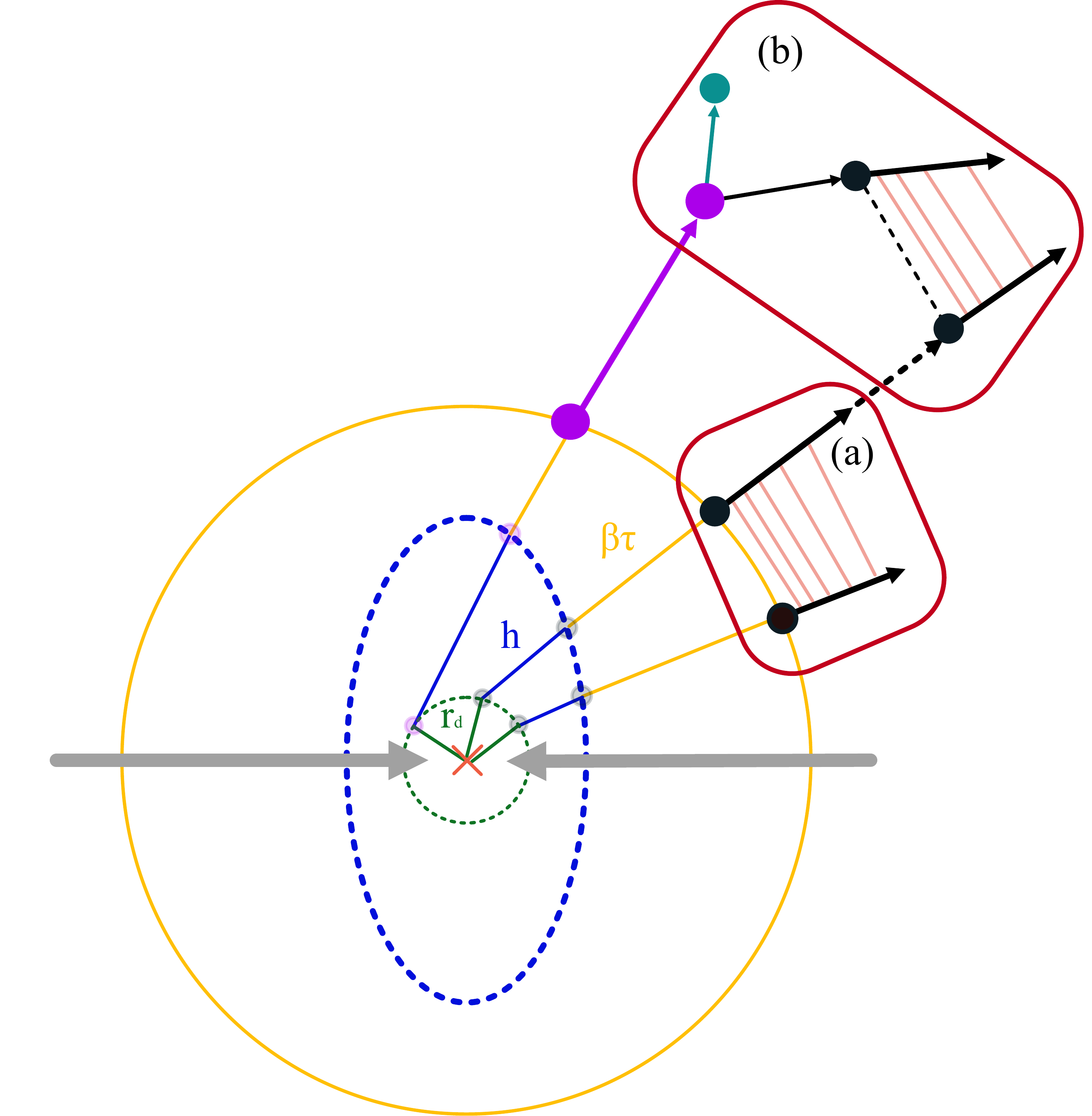}
    \caption[]{Qualitative representation of the CECA dynamics with the main input parameters: $r_d$, $h$ and $\tau$. Panel (a) presents the scenario in which the CECA source is evaluated for a pair of primordial particles. Panel (b) shows the case of a primordial particle being paired with the decay of a strongly decaying resonance. Details in the text.}
    \label{fig:ceca:sketch}
\end{figure}

The CECA model is embedded in a Monte-Carlo framework, and it is based on the emission of single particles, subsequently constructing the two- or many-body source functions. In this work the focus is explicitly on two-body systems. The particles are treated as classical point-like objects, with well-defined position, momentum and time components. The modeling of the source relies on an effective parameterization that allows to correlate the space and momentum components of each particle. This can be used as an effective tool to emulate the \mt scaling of the source function. The generation of particles can be divided into four stages explained in \ref{sec:CECA:disp}--\ref{sec:CECA:reso}. A schematic example is provided in Fig.~\ref{fig:ceca:sketch}.
\subsection{Initial scattering process}\label{sec:CECA:disp}
    The cartoon in Fig.~\ref{fig:ceca:sketch} represents a proton--proton collision, and the red cross marks the geometrical center of the event. 
    It is assumed that the location of the initial scattering process among the constituent partons is randomly displaced by a quantity \rdv (dashed green line in Fig.~\ref{fig:ceca:sketch}) with respect to the center of the event and the orientation of the momenta of the scattered fragments are not correlated to their spatial coordinates. In CECA, this \textit{displacement} \rdv follows a L\'evy $\alpha$-stable distribution\footnote{The L\'evy stable distribution is a generic class of PDFs, for which a linear combination of two independent random variables results in the same distribution. Both the Gaussian and Cauchy distributions are special cases of the L\'evy distribution.} and the expected standard deviation is smaller but comparable to the radius of a proton ($\lesssim0.8~$fm).
\subsection{Particle hadronization}\label{sec:CECA:hadr}
    In the beginning of this stage, the assumption is that the microscopic components are partons. They are not explicitly modeled, however considering only light quarks (u,d,s) the collision system can be assumed to expand with a common local velocity for a fixed amount of time. At the end of the expansion, the hadrons are formed simultaneously and acquire their mass. The space coordinates at which the hadron formation takes place is described with $\rdv+\hv$, where \hv follows the profile of an ellipsoid centered around the displacement point. The hadronization surface is evidenced by the dashed blue line in Fig.~\ref{fig:ceca:sketch}, and effectively parameterized by the \textit{hadronization parameter} \hv within CECA.
 \subsection{Particle emission}\label{sec:CECA:tau}   
    According to the femtoscopic principle, the effective emission time is the moment when the investigated particle pairs experience forces predominantly related to their FSI. This implies that the studied particles have well-defined wave functions, and any possible overlap with other particles produced in the event is negligible. Thus, an extra evolution stage is included in the simulation, during which the particles are propagated and separated from one another. This stage lasts a fixed amount of time ($\tau$), during which the particles move on a straight line with a velocity ($\beta$) based on their momentum and on-shell mass, ignoring any significant shift of the momentum due to re-scattering. By default, $\tau$ is the proper time for each particle. Finally, the particles are considered to be "emitted", and their coordinates are used to build the \textit{primordial} particle emitting source. A pair of primordial particles are illustrated in panel a) of Fig.~\ref{fig:ceca:sketch}, where the solid yellow line represents the core source.
 \subsection{Production through resonances}\label{sec:CECA:reso}   
    Many of the primordial particles are short-lived resonances, that decay after few fm$/c$. However, in this short period the effect of the FSI on the resonances can be ignored, thus the femtoscopic signal is generated from the interaction of the decay daughters with other adjacent particles. An example is presented in panel b) in Fig.~\ref{fig:ceca:sketch}, where the black solid circles represent the investigated particle species, and the purple circles correspond to short-lived resonances. A particle pair can be formed from the decay daughter of a resonance, by pairing it either with a primordial particle, as in b), or to another decay product. On average, these effects lead to an increase of the distance between the paired particles. The femtoscopic formalism demands an equal time of emission between the studied particles, thus CECA monitors the time components of all generated particles, and when the pairing is performed the earlier produced particle is propagated on a straight line until the time of emission of its partner. The correction is applied on the level of pairs in their rest frame. This is illustrated in Fig.~\ref{fig:ceca:sketch}, where the same particle is used to build a pair in a) and b), however in b) it has been propagated to match the time of emission of the decay product to which it is paired. To perform the simulation, the CECA framework requires as an input the amount of primordial and resonance particles and their momentum distributions. In addition, the resonances are characterized by their lifetime, decay channels and branching ratios.

The CECA framework is intended to be an effective model describing existing experimental data with a limited set of observables, which can then be used to test and constrain transport models, such as EPOS, Pythia or AMPT~\cite{EPOS:2013ria,PYTHIA:2006za,AMPT:2004en}, ultimately leading to a better understanding of the microscopic properties of hadronization. The intrinsic limitation is the available data, and thus the parameterization has to be minimalistic in order to avoid overfitting and to provide a unique set of parameters to describe the emission source. To achieve this, it is assumed that the displacement \rdv is Gaussian and identical in all spatial directions, consequently described by a single scalar \rd, representing the standard deviation of the underlying distribution. The ALICE data, to which CECA is applied in this work, has an acceptance in pseudorapidity of $|\eta|<0.8$~\cite{ALICE:Source}. Consequently, the CECA simulation will be more sensitive to the transverse component (\hT) of \hv, thus without loss of generality it is assumed that \hz is zero, and $h=\hT$ describes both the $x$- and $y$- components. By definition, the particle emission is described with a single parameter ($\tau$), thus no further simplifications are required. 

\subsection{The CECA source function}

\begin{figure*}[ht]
    \centering
    \includegraphics[width=0.48\textwidth]{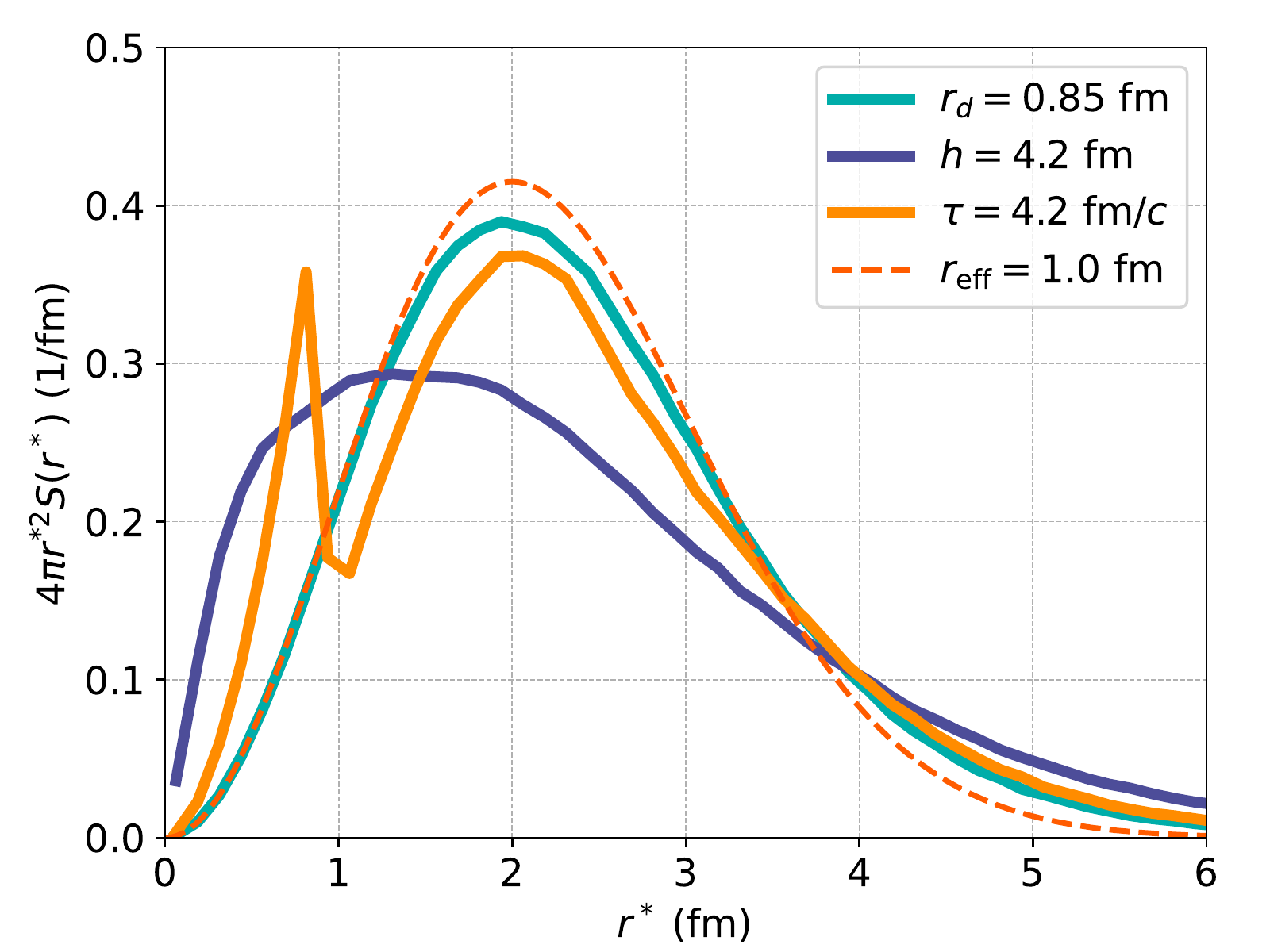}
     \includegraphics[width=0.48\textwidth]{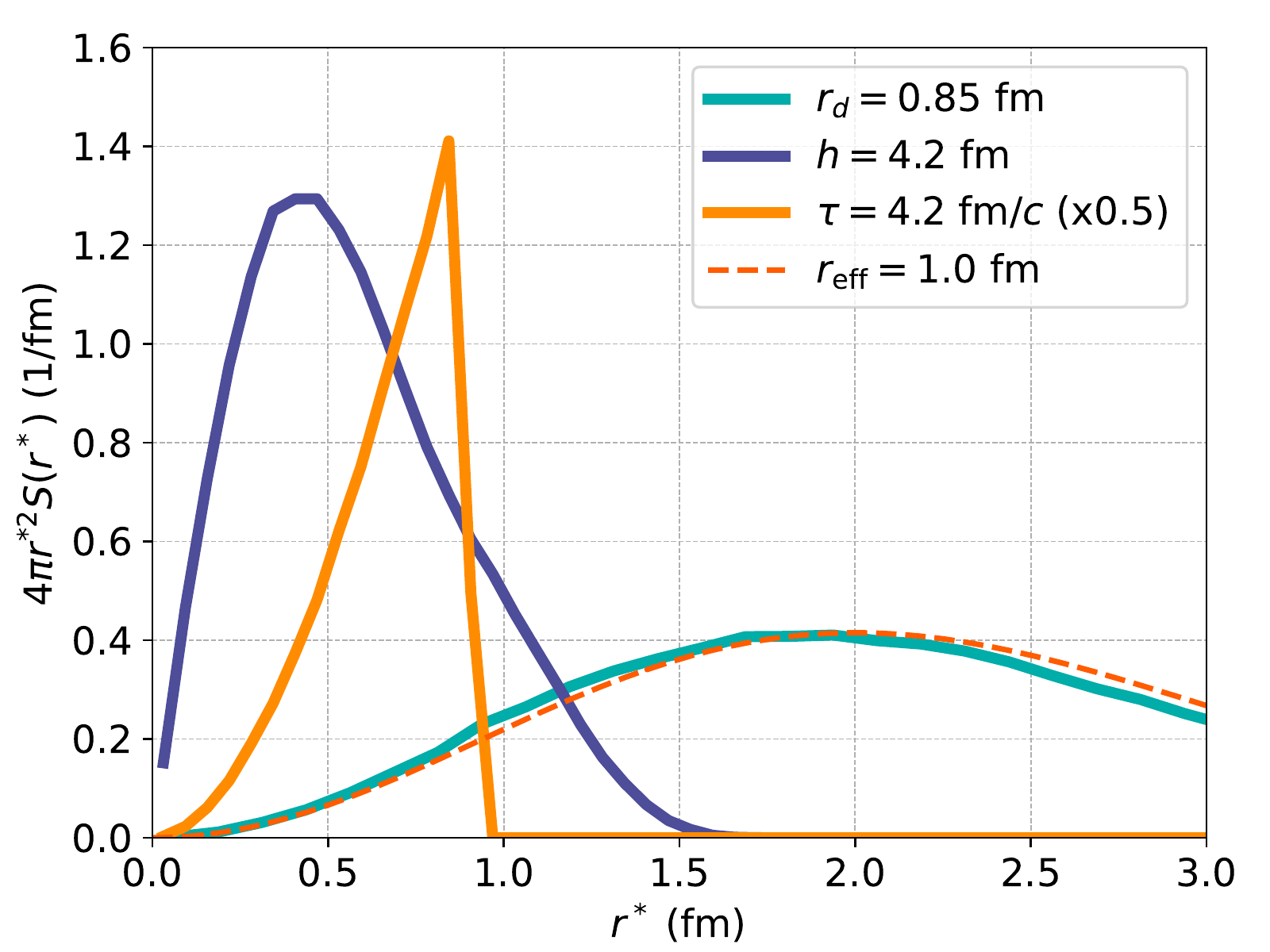}   
    \caption[]{The total source function (left) and the primordial core (right) obtained using only one of the three available parameters. The parameters are chosen to result in an average mean value of the total source similar to that of a Gaussian source (dashed line) of $r_\mathrm{eff}=$1.0~fm.}
    \label{fig:ceca:src}
\end{figure*}

\begin{figure}[ht]
    \centering
    \includegraphics[width=0.45\textwidth]{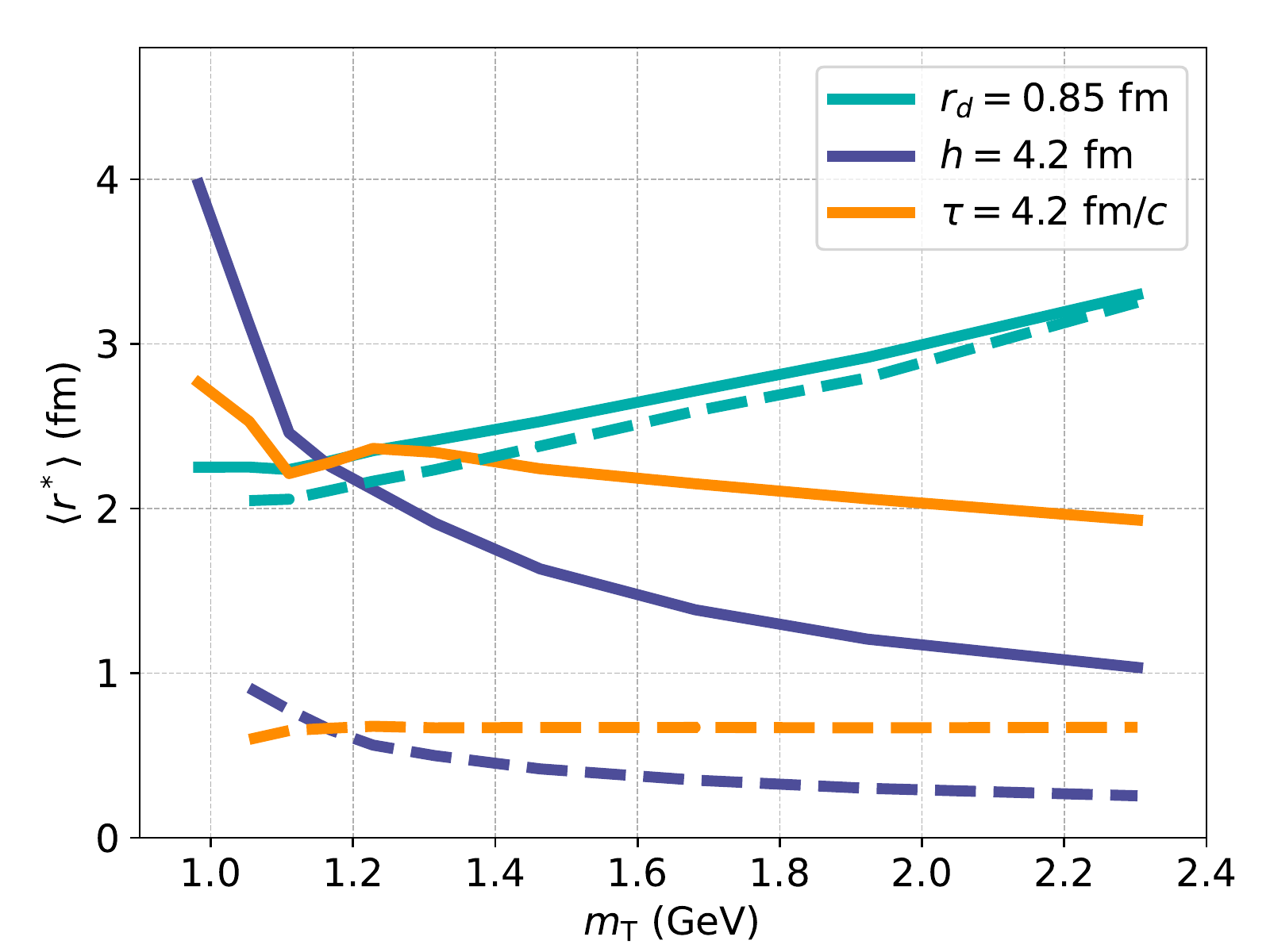}
    \caption[]{The \mt scaling for each of the CECA sources from Fig.~\ref{fig:ceca:src}. The solid lines represent the total source, while the dashed lines are the results for the corresponding primordial (core) sources.}
    \label{fig:ceca:mtEXAMPLE}
\end{figure}

The typical emission source sizes in pp collisions at TeV energies is c.a. $r_\text{eff}\approx1$~fm~\cite{ALICE:Run1}, under the consideration of a Gaussian parameterization of the source function. The corresponding PDF is shown as a dashed line in Fig.~\ref{fig:ceca:src}. To study the effect of each of the ${\rd, h, \tau}$ parameters, the CECA framework has been employed to simulate the source function for the \pP system using three scenarios, in which only one of these parameters is non-zero and its value is tuned to obtain a PDF of similar mean as the Gaussian source. It is highly likely that none of these parameters are capable of providing physically meaningful modeling of the source, thus the next examples are provided with the goal of gaining a qualitative understanding of the influence that each parameter has on the properties of the emission.
The simulation includes both the primary protons and the production through short-lived resonances. 
The momentum distributions of all relevant particles, as well the properties specific to the resonances, are evaluated following the ALICE analysis presented in~\cite{ALICE:Source}. Further details will be provided in Section~\ref{sec:pp_pLCorr}. 
The left panel of Fig.~\ref{fig:ceca:src} contains the CECA PDFs (solid lines) of the total source functions, while the right panel shows the corresponding primordial (core) sources. In both cases the particle pairs used to build the two-body source functions are those of relative momentum $\ks<100$~\MeVc, corresponding to the femtoscopically relevant region. In fact, the CECA framework is capable of studying the \ks dependence of the source function, nevertheless, this is outside the scope of the present work. 
Further, Fig.~\ref{fig:ceca:mtEXAMPLE} shows the corresponding \mt dependence of the mean $\left<r^*\right>$ of the PDFs. 
The solid turquoise line corresponds to the case of a Gaussian displacement $\rd=0.85~$fm. Unsurprisingly, the resulting PDF is approximately Gaussian, nevertheless, there is a slight tail to the distribution linked to the exponential nature of production through short-lived resonances, as well as to the Lorentz boost into the pair rest frame. The effect of the particle production through resonances is not large, as the total and core sources peak at almost the same value. The reason for this is the random orientation of the space and momentum components of the particles, which allows the resonances to travel in a direction opposite to the geometrical center of the event, traversing into the volume close to the collision point and decaying within the primordial region of particle production. This is not a physical scenario, a statement further confirmed by the increasing trend of $\left<r^*\right>(\mt)$ (Fig.~\ref{fig:ceca:mtEXAMPLE}), which is opposite to the experimental observations. 

However, if $\rd$ is set to zero and the source is modeled only with a hadronization parameter $h=4.2$~fm (solid dark blue line), the difference between the total and core sources (Fig.~\ref{fig:ceca:src}) becomes large, while the \mt scaling is inverted and qualitatively matches the experimental observations. This implies a very compact primordial source, with an average distance between the particles comparable to or smaller than the radius of the proton. The PDFs of both the primordial and total source are of a non-Gaussian shape. Invoking only an emission parameter $\tau=4.2~$fm$/c$ (solid orange line) results in a primordial source similarly compact as in the previous case, nevertheless, the corresponding PDF has a significantly different shape, possessing a sharp cut-off around 0.9~fm. This is related to the fact that each particle is propagated for a distance $\beta\cdot\tau\approx\ks\cdot\tau/\sqrt{k^{*2}+m^2}$, under the consideration of $\ks<m$. The sharp cutoff in the source function is observed as $\tau$ is constant and in the present calculation the femtoscopic source is evaluated using the cut-off $\ks<100~$\MeVc, which translates into $\beta\cdot\tau<0.88$~fm. Indeed such a sharp structure is non-physical, nevertheless, if $\tau$ is used in convolution with the displacement and hadronization parameters, the effect will be smeared out.
The \mt scaling related to the emission parameter is flat for the core source and has a slightly decreasing behavior for the total source. This implies that the production of particles through short-lived resonances has a small, yet non-negligible, influence on the \mt scaling.

Summarizing Figs.~\ref{fig:ceca:src} and~\ref{fig:ceca:mtEXAMPLE}, the displacement parameter $\rd$ closely resembles the traditionally used Gaussian emission source, and the resulting \mt scaling has a rising behavior. The hadronization parameter $h$ introduces space-momentum correlations, as all particles effectively propagate (flow) radially away from the collision point, resulting in an \mt scaling that qualitatively matches the experimental observations. The emission parameter $\tau$ results in a non-smooth source function, which is a consequence of the cut-off in \ks when evaluating the source function. The resulting \mt scaling is approximately flat.

%% file: Chapters/Analysis.tex
\section{\texorpdfstring{\pP and \pL correlations}{}}\label{sec:pp_pLCorr}

The ALICE measurement of the \pP and \pL correlations in high-multiplicity \pp collisions at 13~TeV, performed differentially in \mt~\cite{ALICE:Source}, provides a great opportunity to validate the CECA framework and investigate the hypothesis of a common emission source for these two baryons. The \pP interaction is known with great precision due to the large amount of existing scattering data~\cite{Stoks:ppPWA}, which have been used to constrain the properties of the strong potential. In particular, the Argonne $v18$ (AV18) potential provides the possibility to model the \pP system with high accuracy at low \ks, accounting for s-, p- and d-waves~\cite{Mihaylov:2018rva,Wiringa:AV18}. 

The \pL interaction, on the other hand, is not as well-known due to the scarce data for this system~\cite{Alexander:pLambda,Sechi-Zorn:pLambda,Eisele:pLambda}. The $\chi$EFT, evaluated to \textit{next-to-leading order} (NLO)~\cite{Haidenbauer:NLO13,Haidenbauer:NLO19} and recently to \textit{next-to-next-leading-order} (N$^2$LO)~\cite{Haidenbauer:NNLO}, represents the state-of-the-art tool to study the \pL interaction. However, the output of $\chi$EFT calculations depends on the determination of the so-called low-energy constants, which need to be constrained by experimental data. The existing scattering data on \pL have limited statistical significance, particularly at low energies. Additional experimental constraints on the \pL system come from the measurements of hypernuclei binding energies, whose interpretation is model-dependent and includes many-body effects~\cite{Hashimoto:pL_hyper,Gal:pL_hyper}.
Another recent analysis by the ALICE collaboration delivered the \mt-integrated measurement of the \pL correlation function in high-multiplicity pp collisions, in which the statistical precision improves by more than one order of magnitude compared to the scattering data and extends the range of accessible momenta practically down to 0 MeV~\cite{ALICE:pL}. By constraining the emission source to \pP correlations using the RSM, it has been shown that the NLO potential deviates by around $3\sigma$ from the data, suggesting either a weaker two-body attraction between the proton and the \Lam or a larger emission source.

In this work the analyzed \pP and \pL correlations are split into, respectively, 7 and 6 differential bins of \mt~\cite{ALICE:Source}. In the associated publication of the data, the RSM has been successfully applied independently in each range of \mt, nevertheless, the model cannot describe all bins simultaneously with a common set of parameters. This is addressed in the present work, as the adaptation of CECA into the CATS framework allows to fit all 13 correlation functions simultaneously, using a common source for protons and \Lam baryons described by a single set of ${\rd, h, \tau}$ parameters. Two scenarios are used for the fit, in which the \pL interaction is treated differently.

\subsection{\pL interaction}\label{sec:pp_pLCorr:pL}
\begin{figure}[h!]
    \centering
    \includegraphics[width=0.45\textwidth]{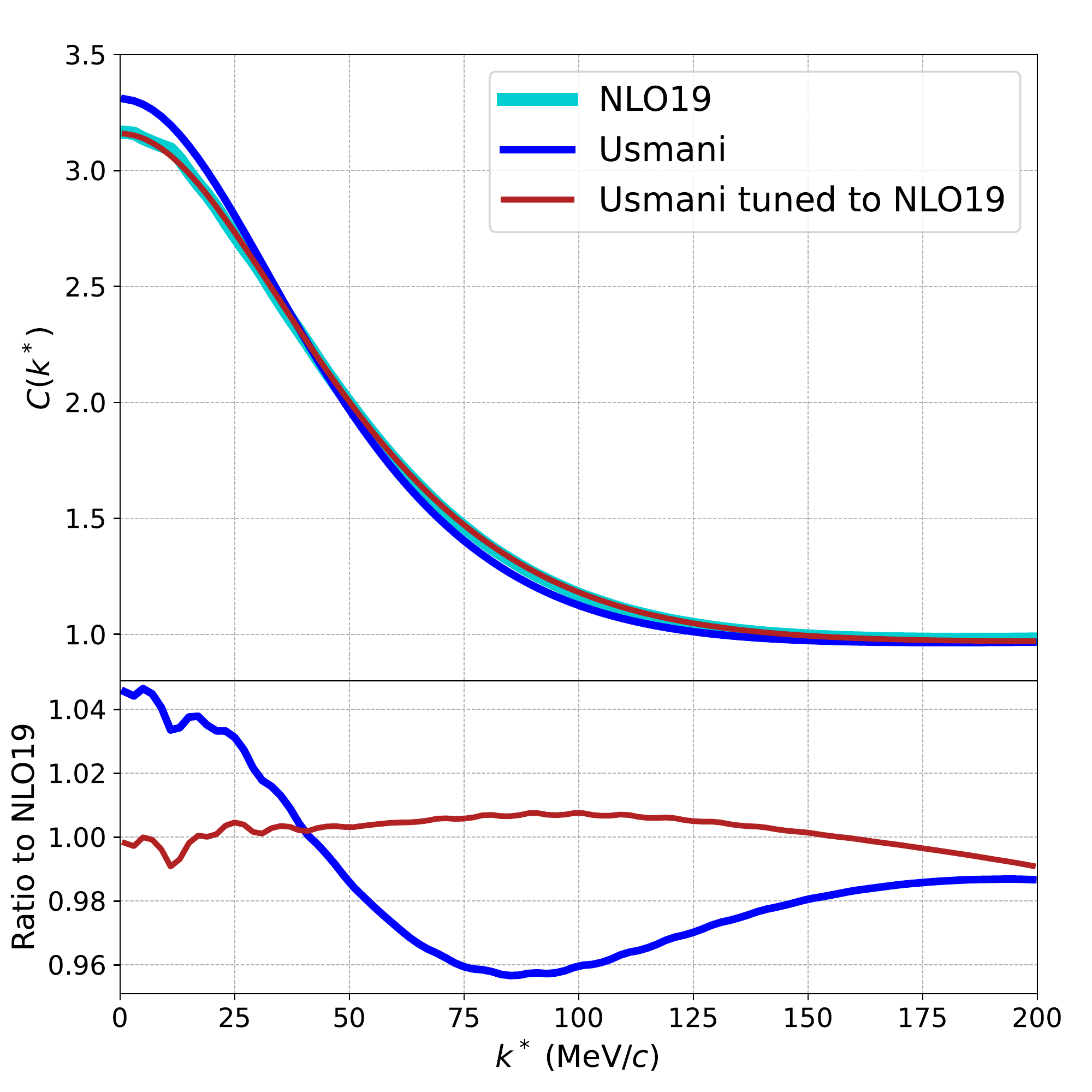}
    \caption[]{The genuine \pL correlation function evaluated using CATS for a Gaussian source of 1.2~fm. The turquoise line corresponds to the NLO19 calculation, which includes s- and d-waves, as well as the coupling to \NS~\cite{Haidenbauer:NLO13,Haidenbauer:NLO19}. The blue line is evaluated using the standard parameterization of the Usmani potential, in which only s-waves are considered. The dark red line is Usmani parameterization, fine tuned to NLO19. The bottom panel shows the ratio to the original NLO19.}
    \label{fig:ana:usm_vs_nlo}
\end{figure}

The genuine \pL term in the total correlation function in Eq.~\ref{eq:LamPar} is modeled assuming at first the NLO19 chiral potentials, following the work in~\cite{ALICE:pL}.
The NLO19 potential delivers a weaker attraction in the triplet state with respect to previous tunings and the cut-off parameter $\Lambda_\mathrm{cut}=600$~MeV, needed for the regularization of the potential in the Lippman-Schwinger equation~\cite{Haidenbauer:NLO13}, delivers the best description for both femtoscopic~\cite{ALICE:pL} and scattering data~\cite{Haidenbauer:NLO19}. The scattering length in the S=0 channel is $f_0=2.91~$fm\footnote{In this work, the convention of a positive scattering length for an attractive interaction is used.}, while in the S=1 channel the value is $f_1=1.41~$fm. 
Additional details on the NLO19 potential can be found in~\cite{Haidenbauer:NLO19}.
In Fig.~\ref{fig:ana:usm_vs_nlo}, the genuine correlation function obtained assuming the NLO19 \pL interaction is shown (turquoise). A Gaussian source size of 1.2~fm is assumed. The corresponding correlation function for an underlying interaction described by the Usmani potential~\cite{Bodmer:1984gc} is shown in blue. The scattering lengths of the $\chi$EFT and the Usmani parameterizations are summarized in Tab.~\ref{tab:ScatPars}.
The two approaches have similar scattering parameters in the singlet channel, but the Usmani potential results in a larger attraction in the triplet channel, yielding an enhanced correlation function. The Usmani potential is composed of a repulsive central term (repulsive core), depending only on the distance $r^*$ between the two particles, and of a two-pion exchange contribution multiplying both the attractive spin-independent part and the repulsive spin interaction~\cite{Bodmer:1984gc}.
The repulsive core term $V_C$ of the Usmani interaction is a dominant contribution at short inter-particle distances and hence represents an important component of the correlation signal. The parameterization of $V_C$ is given by a Woods–Saxon function
\begin{equation}\label{eq:UsmCore}
    V_C(r^*) = W_C\left[1+\mathrm{exp}\left(\frac{r^*-R_C}{d_C}\right)\right]^{-1},
\end{equation}
where the individual parameters are determined empirically, anchoring them to available scattering data. The default values are $W_C=2137$ MeV, $R_C=0.5$ fm, $d_C=0.2$ fm and deliver the blue line in Fig.~\ref{fig:ana:usm_vs_nlo}. 
Fine-tuning the S=1 parameters ($W_C=2279~\mathrm{MeV}$, $R_C=0.3394~\mathrm{fm}$, $d_C=0.2614~\mathrm{fm}$) while leaving the S=0 channel unchanged reproduces the NLO19 value of $f_1=1.41~$fm. The resulting spin-averaged \pL correlation functions for the fine-tuned Usmani potential and the original chiral NLO19 calculation are in good agreement for $\ks$ values below 180~\MeVc, as shown in Fig.~\ref{fig:ana:usm_vs_nlo}. 

This result implies that the Usmani potential and $\chi$EFT can be used interchangeably to model the \pL correlation function at low \ks. In the present work, the Usmani potential will be used to study the interaction strength of the \pL system, by fitting the $W_C$, $R_C$ and $d_C$ parameters to the high-precision ALICE data published in~\cite{ALICE:Source}. The correlation function is known to be insensitive to the effects of the individual spin channels, for this reason, without loss of generality, only the parameters of the S=1 channel will be varied. The choice to use explicitly the S=1 state is motivated by two factors. First, this state is dominant within the correlation function, as it contributes with a weight of 3/4 (Eq.~\ref{eq:Spin}). Second, the binding energy of the hypertriton depends on the scattering length of the S=0 state, thus a significant variation of $f_0$ cannot be performed without studying the subsequent effect on the hypertriton, which is outside the scope of the present work. 
\begin{table}
\centering
\begin{tabular}{c|c|c}
\hline
   & $f_0~$(fm) & $f_1~$(fm) \\
\hline
 NLO19  & 2.91 & 1.41 \\
\hline
 N$^2$LO  & 2.80 & 1.56 \\
\hline
 Usmani  & 2.88 & 1.66 \\
\hline
 Usmani (NLO19)  & 2.88 & 1.41 \\
\hline
\end{tabular}
\caption[]{The scattering length of the singlet (left column) and triplet (right column) \pL channel for different theoretical models.}\label{tab:ScatPars}
\end{table}

\subsection{Analysis details}\label{sec:AnaDetails}

The correlation functions in the present work have been measured by the ALICE collaboration in \pp collisions triggered for high-multiplicity at 13~TeV~\cite{ALICE:Source}. The \pP data is split into 7 \mt bins, and the \pL into 6 bins. The analysis procedure performed by ALICE is closely followed, and the main steps are highlighted below. 

The measured correlation functions are obtained by constructing the ratio of the same- and mixed-event samples (\SEks and \MEks). These distributions are not corrected for any experimental effects, most notably the acceptance, misidentifications, feed-down from weakly decaying particles, and momentum resolution~\cite{ALICE:Run1}. All of these effects have to be included in the modeling of the data, which is achieved using the CATS framework. To evaluate the correlation function, CATS requires either an interaction potential or a wave function, describing a specific (S,L) state. The source function can be modeled either by an analytic function, such as a Gaussian source, or linked to a more advanced framework, such as CECA. The total correlation function provided by CATS is
\begin{equation}\label{eq:CkCats}
\begin{split}
        C_\mathrm{CATS}(\ks) = & \mathcal{N}\cdot \int dk^*_\mathrm{true} \mathcal{M}(\ks,k^*_\mathrm{true})\cdot \\
    & \sum_i \int dk^*_\mathrm{mother} \lambda_i \mathcal{T}_i(k^*_\mathrm{true},k^*_\mathrm{mother}) C_{i}(k^*_\mathrm{mother}),
\end{split}
\end{equation}
where $\lambda_i$ are the weights corresponding to each possible contribution $i$ to the correlation signal (genuine, feed-down, etc.), $\mathcal{M}$ is a momentum smearing matrix that transforms the "true" relative momentum ($k^*_\mathrm{true}$) to the measured one (\ks), $\mathcal{T}_i$ is a transformation matrix for the residual feed-down correlations and $\mathcal{N}$ is a normalization constant. 
The residual contributions to the correlation signal arise when a measured pair consists of at least one particle that is the decay product of a long-lived ($c\tau\gg$fm) resonance. In that case, the FSI signal is not related to the measured pair, but to the interaction between the mother particles. For example, a \pL pair which consists of a primary proton and a secondary \Lam, stemming from the decay of a $\rm\Xi^-$ particle, will carry a residual signal from the FSI of the original \pXim pair~\cite{ALICE:Run1}. The true \pL relative momentum $k^*_\mathrm{true}$ will differ from the original $k^*_\mathrm{mother}$ of the \pXim pair, thus requiring the correction by $\mathcal{T}_i(k^*_\mathrm{true},k^*_\mathrm{mother})$. In the case of primary or misidentified pairs this correction is not required and the matrix becomes equal to the identity operator. Following the ALICE analysis, in the present work the considered (non-flat) feed-down channels are $\pL\rightarrow\pP$, $\pXim\rightarrow\pL$, $\pXio\rightarrow\pL$ and $\pSo\rightarrow\pL$. The corresponding $\lambda$ parameters are provided in~\cite{ALICE:pL}, and varied within their uncertainties. The correlation functions of the $\pXi$ interaction are based on lattice calculations~\cite{Sasaki:pXi} and the \pSo on $\chi$EFT~\cite{Haidenbauer:NLO13}. The \pL experimental correlation functions are fitted and the corresponding results are used as a direct input for the $\pL\rightarrow\pP$ residual contributions. The \pL interaction is modeled using the Usmani potential with the repulsive core of the triplet channel either fixed to reproduce the scattering length of the $\chi$EFT NLO19 or fitted (see Section~\ref{sec:pp_pLCorr:pL}). Finally, the experimental measurements may contain non-femtoscopic correlations, which are typically broad and smooth structures that can be described by a polynomial function~\cite{ALICE:Run1}. In the region $\ks<200$~\MeVc the effect of these correlations is expected to be minor and approximately flat. In the present analysis the correlation functions are fitted up to $\ks<180$~\MeVc, allowing the effective modeling of the non-femtoscopic correlations by the constant $\mathcal{N}$. This parameter is fitted independently for each of the 13 correlation functions. 
The modeling of the source with CECA requires the single-particle spectra of the measured protons and \Lam particles. Ideally, these should account for the kinematics of particle production through resonances and include a full-scale simulation within the acceptance and efficiency of the detector. Such a dedicated analysis is not possible without detailed knowledge of the performance of ALICE, thus in the present work a simplified approach is adopted, in which the \pt distributions of the proton and \Lam baryons are taken from the highest multiplicity class of the fully corrected ALICE spectra~\cite{ALICE:ppSpectra,ALICE:pLambdaSpectra}. 
The short-lived resonances feeding into protons and \Lam baryons are assumed to have similar momentum distributions. Further, the published ALICE data have an acceptance for the protons $\pt=[0.5, 4.05)~$\GeVc~\cite{ALICE:Source}, which has been enforced in the present analysis. To sample in 3 dimensions, the pseudorapidity distribution of both protons and \Lam baryons is assumed flat within the acceptance of $\left|\eta\right|<0.8$, while $\varphi$ is assumed uniform. 
Following the procedure within the ALICE analysis~\cite{ALICE:Source}, the properties of the resonances are averaged, and the feed-down to each of the proton and \Lam species is modeled by a single effective resonance. According to the statistical hadronization model, the average mass of the resonances decaying into protons (\Lam baryons) is 1.36 (1.46)~\GeVcc and the average lifetimes are 1.65 (4.69)~fm/$c$~\cite{ALICE:Source,Becattini:2009ee}.

Summarizing the analysis procedure, Eq.~\ref{eq:CkCats} is used to fit the measured \pP and \pL correlations in each \mt bin. All relevant genuine and residual correlation functions are computed from the Koonin-Pratt equation (\ref{eq:KooninPratt_Simple}) using the CATS framework. The interaction potentials are fixed for \pP and all feed-down contributions, while the \pL interaction is treated separately, as discussed in Section~\ref{sec:pp_pLCorr:pL}. The source function is evaluated using the CECA framework and is common for all correlation functions and parameterized using the 3 variables ${\rd, h, \tau}$. Additionally, each correlation function is independently renormalized by a free parameter $\mathcal{N}$. The $\lambda$ parameters and momentum smearing matrices required by Eq.~\ref{eq:CkCats} are identical to the \mt integrated ALICE analysis~\cite{ALICE:pL}.

\subsection{Results and discussion}
\begin{figure*}[ht]
    \centering
    \includegraphics[width=0.46\textwidth]{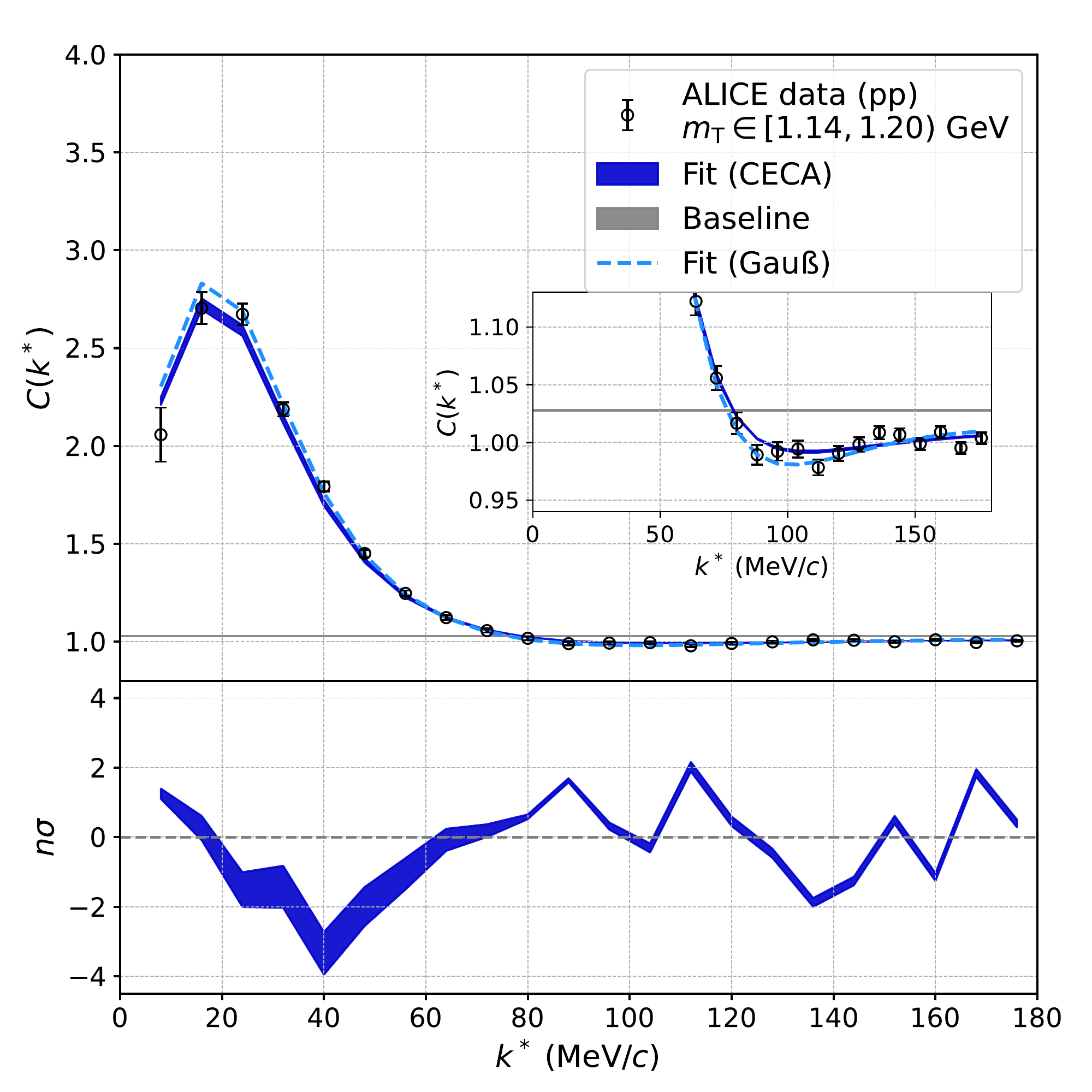}
    \includegraphics[width=0.46\textwidth]{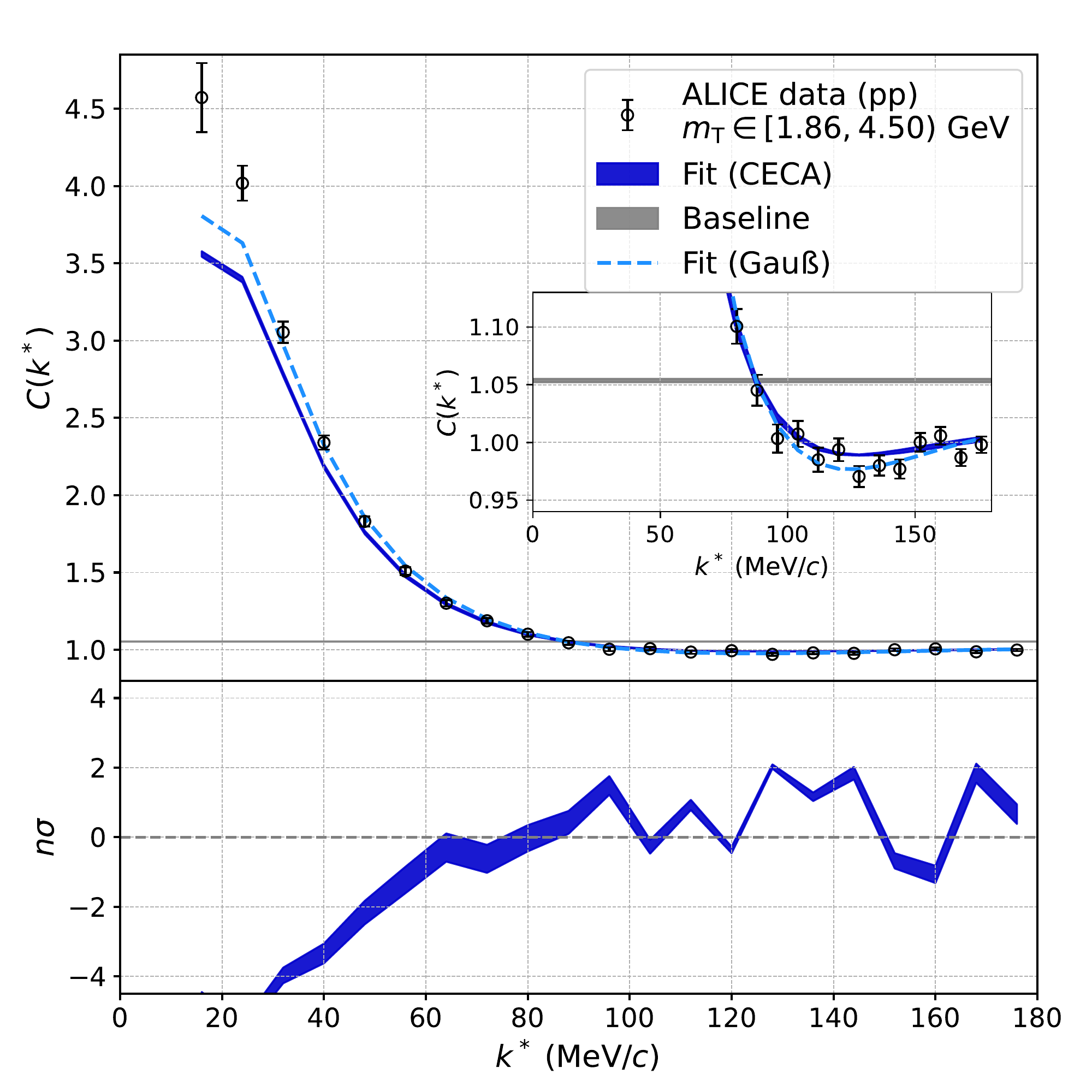}\\
    \includegraphics[width=0.46\textwidth]{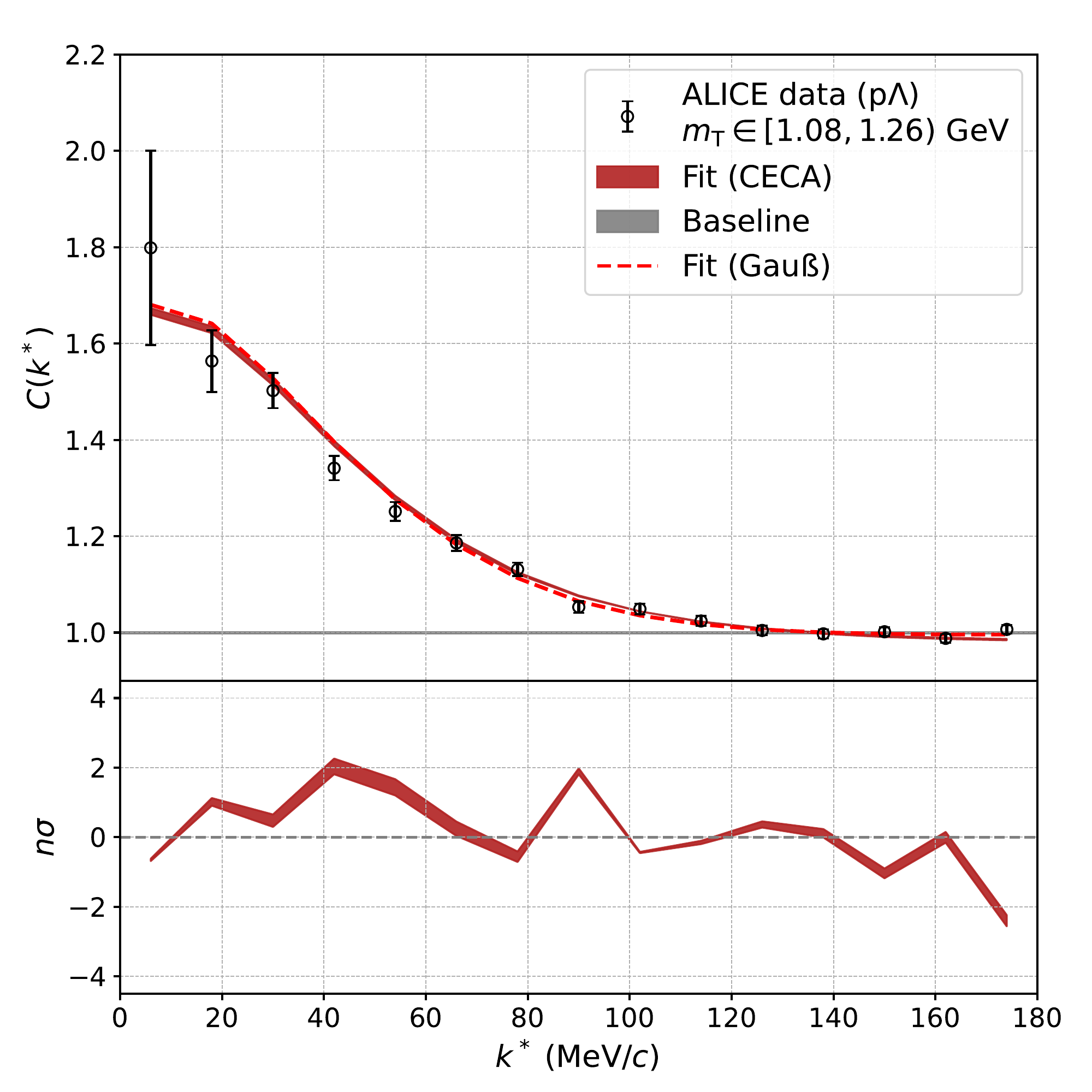}
    \includegraphics[width=0.46\textwidth]{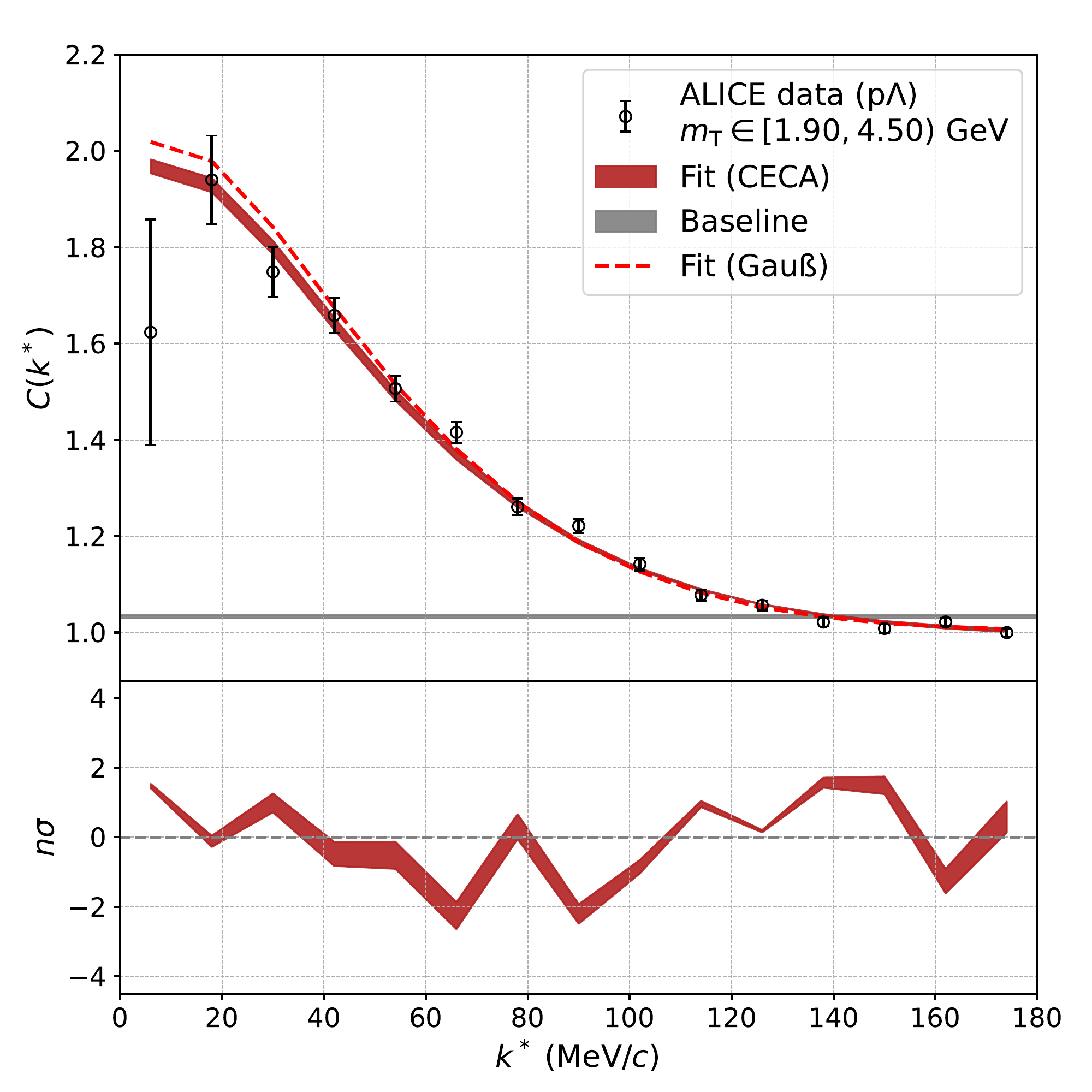}
    \caption[]{The \pP (top) and \pL (bottom) correlation functions in one bin of low \mt (left) and one bin of large \mt (right). The \pL interaction is modeled by the Usmani potential fine-tuned to $\chi$EFT (NLO19). The data (black circles with error bars) are fitted by using the CECA source (bands), and the result using a Gaussian fit is plotted as a reference (dashed lines). The baseline corresponds to the non-femtoscopic contributions to the fit, modeled by $\mathcal{N}$ (see Eq.~\ref{eq:CkCats}). The bottom sub-panel of each plot represents the deviation, in number of $\sigma$, of the fit to the data.}
    \label{fig:ana:Ck_pp}
\end{figure*}

Following the analysis procedure described in Section~\ref{sec:AnaDetails}, the 7 (6) \mt bins of the \pP (\pL) correlations have been simultaneously fitted, modeling the \pL using the Usmani potential tuned to the NLO19 scattering parameters. Figure~\ref{fig:ana:Ck_pp} shows the resulting correlation functions in one bin of low \mt (left) and one bin of large \mt (right). The black points represent the ALICE data, and the error bars are the square root of the quadratic sum of the statistical and systematic uncertainty. The blue (dark red) bands show the \pP (\pL) fit results from CATS, using the CECA source, including the uncertainties related to the free fit parameters and to the $\lambda$ parameters. The dashed lines are the fit results under the assumption of a Gaussian source, where each correlation function is fitted separately. The resulting values for $r_\mathrm{eff}$ are compatible with the published ALICE results~\cite{ALICE:Source}.
\begin{table}
\centering
\begin{tabular}{c|c|c|c}
\hline
   & Usmani & Usmani (NLO19) & Usmani (Fit) \\
\hline
$\chi^2$ & - & 473 & 371 \\
\hline
$d~$(fm) & - & $0.288\pm0.013$ & $0.176\pm0.005$ \\
\hline
$h_\mathrm{T}~$(fm) & - & $3.23^{+0.05}_{-0.30}$ & $2.68^{+0.06}_{-0.04}$ \\
\hline
$\tau~$(fm/$c$)& - & $3.26^{+0.16}_{-0.04}$ & $3.76^{+0.05}_{-0.03}$ \\
\hline
$f_0~$(fm) & 2.88 & 2.88 & 2.88 \\
\hline
$f_1~$(fm) & 1.66 & 1.41 & $1.15\pm0.07$ \\
\hline
$W_C~$(MeV) & 2137 & 2279 & $2332^{+2}_{-39}$ \\
\hline
$R_C~$(fm) & 0.5 & 0.3394 & $0.3455\pm0.0010$ \\
\hline
$d_C~$(fm) & 0.2 & 0.2614 & $0.2575^{+0.0020}_{-0.0000}$\\
\hline
\end{tabular}
\caption[]{Summary of the fit results to the femtoscopic data. The first column is the default Usmani parameterization. The second column is the case of fine-tuning the Usmani potential to reproduce the $f_1$ of $\chi$EFT (NLO19). The third column is the result from performing a fit without constraining the Usmani repulsive core. The $\chi^2$ is evaluated from 243 number of data points.}\label{tab:parameters}
\end{table}

It is observed that the fits perform with similar precision regardless of the source function. Unlike the Gaussian source fit, the CECA model has only 3 common parameters to describe all 13 correlation functions and has an intrinsic \mt dependence. The resulting parameters are summarized in Tab.~\ref{tab:parameters}. The \rd value of around 0.3~fm is smaller than the radius of the proton, which is compatible with the assumption that the dominant factor influencing this parameter is the fluctuation of the position of the initial scattering process between the partons of the colliding beams. The large hadronization parameter $h\sim3~$fm highlights the strong spatial-momentum correlations, pointing to a collective behavior of the system, most likely linked to a radial expansion before the hadronization takes place. An interesting observation is the necessity of a large $\tau\sim3~$fm$/c$, which points to a phase within the formation of the emission source, in which the particles already hadronized and acquired their on-shell mass, but did not decouple from the expanding system. This can be associated with a short re-scattering phase. Nevertheless, a microscopic explanation is not possible without the use of more sophisticated transport models, such as AMPT, EPOS or Pythia~\cite{AMPT:2004en,EPOS:2013ria,PYTHIA:2006za}. 

\begin{figure*}[ht]
    \centering
    \includegraphics[width=0.46\textwidth]{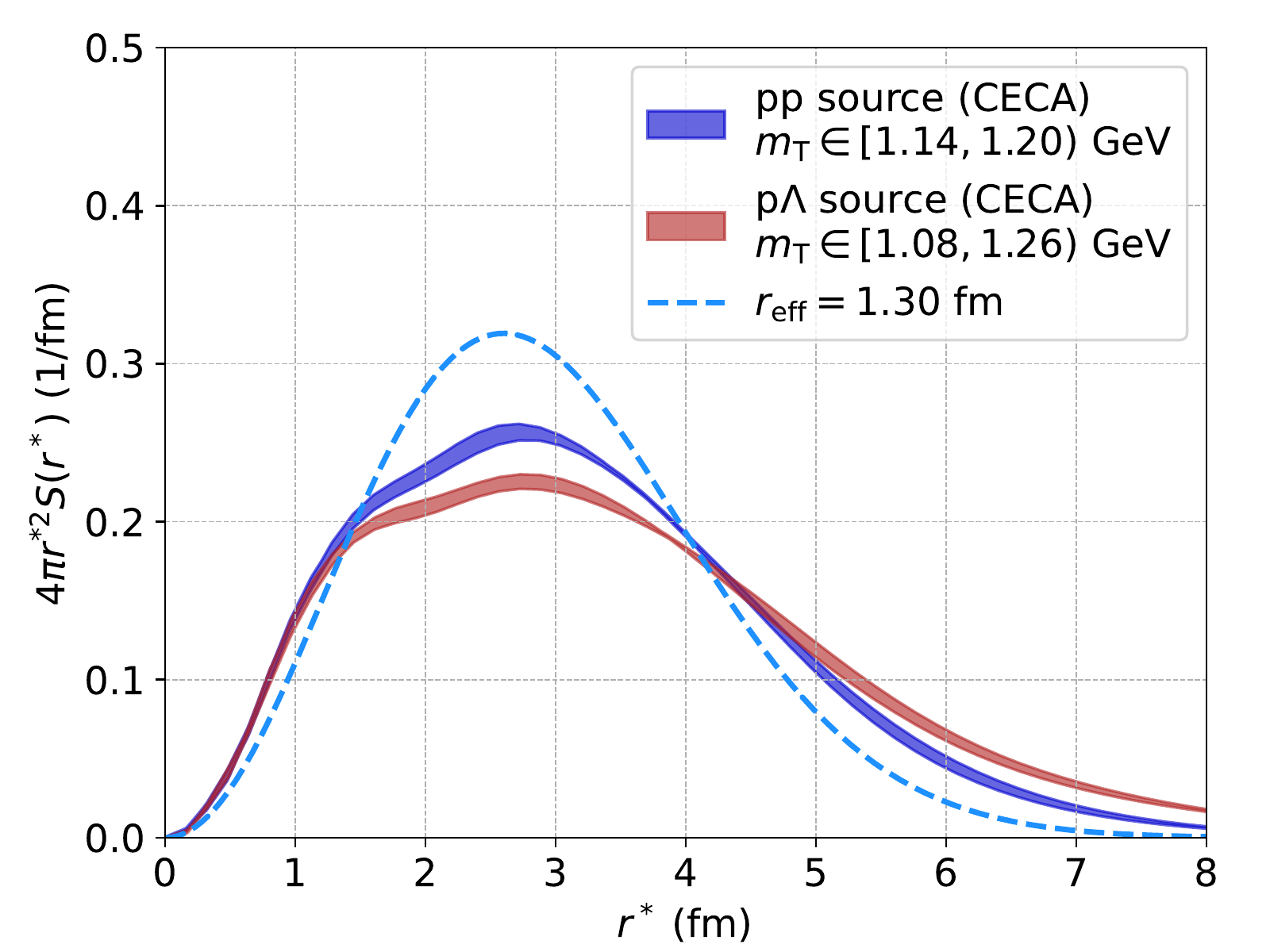}
    \includegraphics[width=0.46\textwidth]{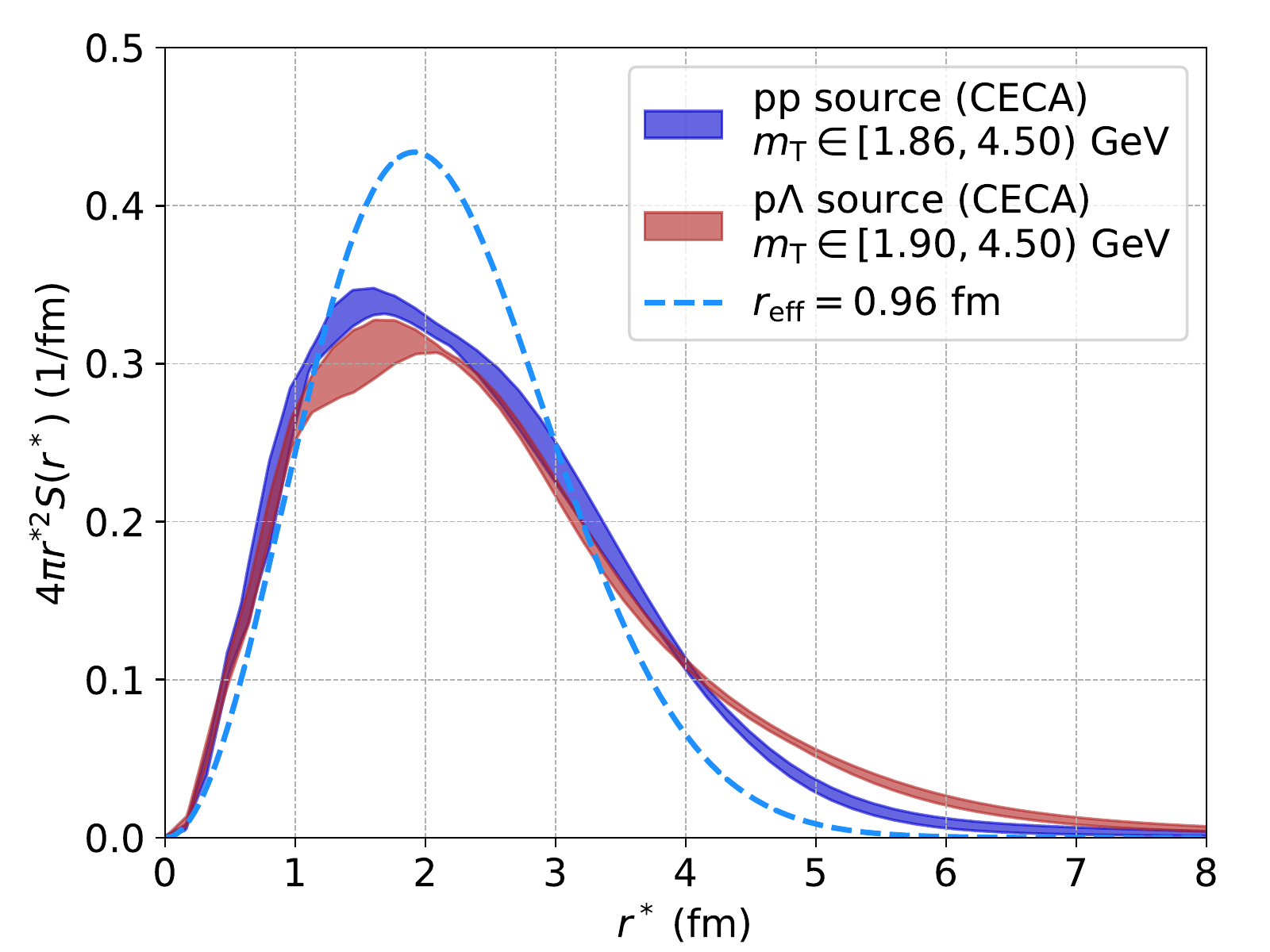}
    \caption[]{The \pP (blue) and \pL (dark red) total source functions in one bin of low \mt (left) and one bin of large \mt (right). The \pL interaction is modeled by the Usmani potential fine-tuned to $\chi$EFT (NLO19). The dashed line corresponds to the effective \pP Gaussian source ($r_\mathrm{eff}$) extracted by the ALICE collaboration, by analyzing the same data~\cite{ALICE:Source}.}
    \label{fig:ana:Sr_pp}
\end{figure*}
\begin{figure}[ht]
    \centering
    \includegraphics[width=0.46\textwidth]{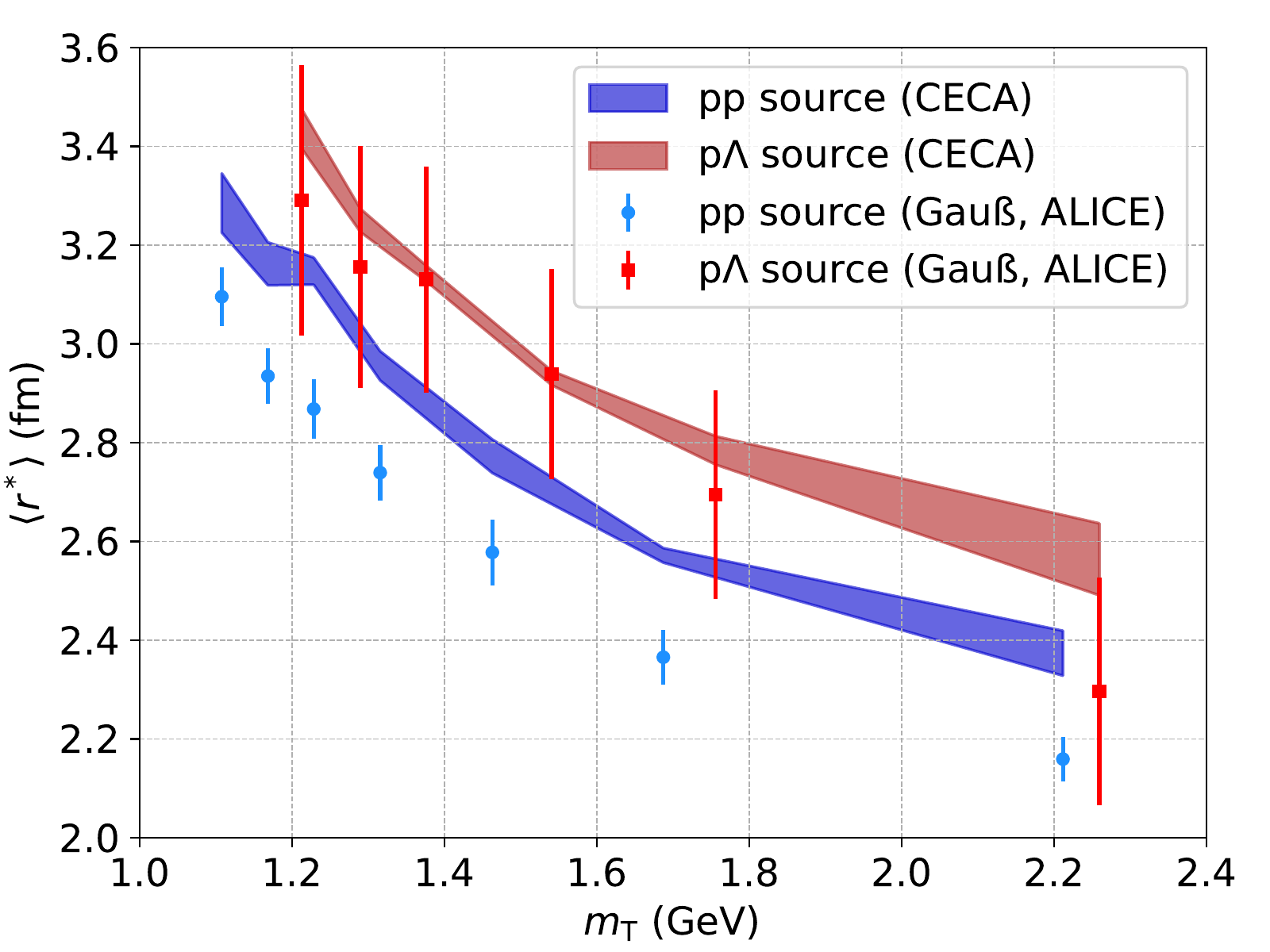}
    \caption[]{The \mt scaling of the source for \pP (blue) and \pL (dark red), obtained by CECA within the current work (bands) and by ALICE (symbols with error bars) within~\cite{ALICE:Source}. In both cases the \pP is modeled using the AV18 potential. The present analysis describes the \pL interaction using the Usmani potential fine-tuned to $\chi$EFT (NLO19), while in~\cite{ALICE:Source} multiple parameterizations of $\chi$EFT (LO and NLO13) are used.}
    \label{fig:ana:mT_def}
\end{figure}

Figure~\ref{fig:ana:Sr_pp} shows the PDFs of the source functions, obtained from CECA, for \pP (blue bands) and \pL (dark red bands) in two different \mt regions. The dashed lines show the corresponding Gaussian sources for \pP. Evidently, the CECA source is non-Gaussian, and has a large tail related to particle production through short-lived resonances. The tail is more pronounced for \pL, due to the larger average lifetime of the associated decay mothers. The peak region is fairly wide, and a careful examination revealed that this is the result of a multi-structured PDF, where the most noticeable differentiation is between purely primordial (core) pairs, which peak at around 1~fm, while the larger \rs values are dominated by the production of pairs in which at least one particle stems from a resonance. 
Figure~\ref{fig:ana:mT_def} summarizes the resulting mean values of the \pP (blue bands) and \pL (dark red bands) emission sources, as a function of \mt. The results are compared to the ALICE analysis, by extracting the mean value from the Gaussian PDF corresponding to the effective source size published in~\cite{ALICE:Source}. The CECA source produces the expected \mt scaling, nevertheless, the mean of the source size is slightly different for \pP. These two results stem from the same data, thus the observed difference can only be due to the non-Gaussianity of the CECA source. This highlights the fact, that the correlation function is sensitive to the shape of the source function, and the latter cannot be described by a single parameter, such as the mean. The $\chi^2$ per \textit{number of data points} (NDP) of the global fit is 473/243, pointing out the need for further improvement. An obvious issue is present in the two largest \mt bins of the \pP correlation functions (Figs.~\ref{fig:ana:Ck_pp},~\ref{fig:ana:Ck_pp_Usm_ALL}), where a significant deviation at low \ks is present, independently on the choice of source function. Interestingly, such a discrepancy is not observed in the \pL correlations. Thus, one possible explanation is that the issue is related to the accuracy of the AV18 potential, which may break down at the very small distances realized at large \mt. Nevertheless, it is not possible to exclude alternative interpretations related to the source function. For example, primordial pairs produced in mini-jets will have a large \mt and will be located very close in space. While pairwise baryon production in mini-jets is not observed in correlation studies~\cite{ALICE:Run1}, it may become relevant at large transverse momentum of the particles. Such an effect will be stronger for \pP compared to \pL of equivalent \mt, due to the lower mass and lack of strange quarks in the system. Ultimately, this will lead to a second effective source of proton pairs at large \mt, which may distort the shape of the overall PDF.

\begin{figure}[h]
    \centering
    \includegraphics[width=0.46\textwidth]{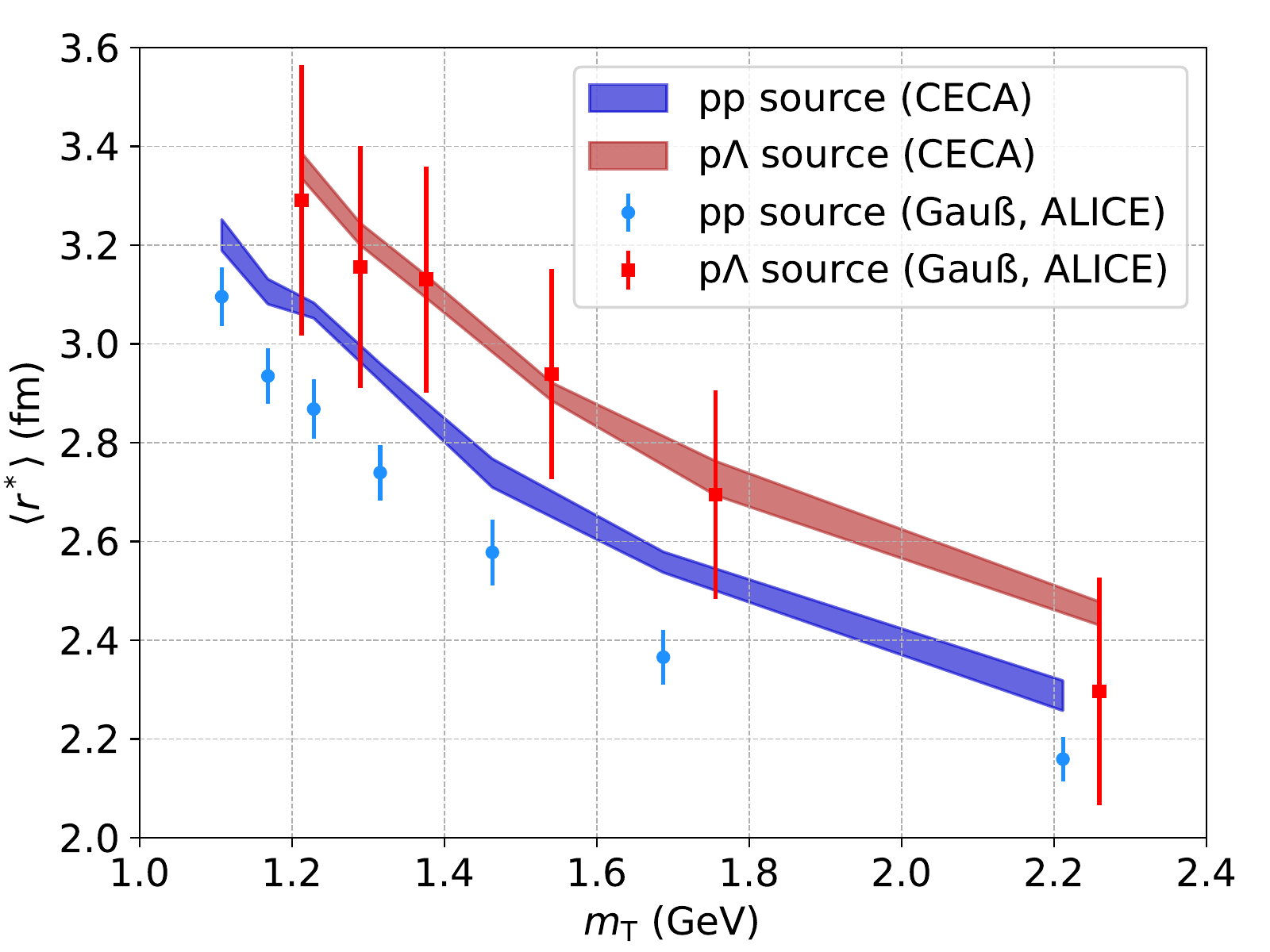}
    \caption[]{The \mt scaling of the source for \pP (blue) and \pL (dark red), obtained by CECA within the current work (bands) and by ALICE (symbols with error bars) within~\cite{ALICE:Source}. In both cases the \pP is modeled using the AV18 potential. The present analysis describes the \pL interaction using the Usmani potential fitted to the femtosopic data, while in~\cite{ALICE:Source} multiple parameterizations of $\chi$EFT (LO and NLO13) are used.}
    \label{fig:ana:mT_fit_Usm}
\end{figure}

As discussed, the low-energy constants of the chiral theory, used to describe the \pL interaction, are only constrained by scattering data, which are of insufficient quality to provide a unique parameterization to the theory. Thus, it is interesting to investigate how a modified \pL interaction influences the fit results. This is achieved by leaving the 3 parameters of the repulsive core of the Usmani potential (Eq.~\ref{eq:UsmCore}) free within the fit procedure, while the source is common for protons and \Lam baryons and modeled by CECA. The resulting \mt scaling, fit results and source distributions are presented in Figs.~\ref{fig:ana:mT_fit_Usm}--\ref{fig:ana:Sr_pp_Usm_ALL}. Summary of the obtained parameters is provided in Tab.~\ref{tab:parameters}. The $\chi^2$ is reduced from 473 to 371, which equals a significance of 9.1$\sigma$, and the resulting scattering length for the \pL triplet channel is $f_1=1.15\pm0.07~$fm. This implies that the assumption of a common emission source for protons and \Lam baryons requires a reduced strength of the \pL attraction, compared to the currently adopted values of 1.41~fm (1.56~fm) by the chiral NLO (N$^2$LO). The alternative explanation is the absence of a common emission source in small collision systems, but this scenario is unlikely, as will be discussed below. To test this hypothesis, the NLO19 scattering length is again adopted to model the \pL interaction, however, the \pP and \pL correlation functions are now fitted independently from one another, using two different sets of source parameters. The resulting combined $\chi^2$ is 381, which is compatible within 2$\sigma$ with the fit result for reduced \pL interaction strength and a common source. Nevertheless, this is only achieved by increasing the \pL hadronization parameter $h$ by 0.7~fm, suggesting a different hadronization time for the \Lam baryon. 
This possibility has been studied by employing a toy model based on CECA, in which the \Lam particles are allowed to either hadronize earlier or later than the proton and examine the effect on $h$. A slightly later \Lam production can mimic a larger $h$, however, this is incompatible with the mass hierarchy of the quarks and the subsequent earlier hadronization times of heavy flavour~\cite{Bellwied:2013cta}. On the other hand, an earlier production can increase the effective \pL source size only if the \Lam baryons are produced at least 1~fm/$c$ earlier than the protons. Presently, there are neither experimental observations nor theoretical predictions to support such a hypothesis, consequently, the observed deviation is likely connected to the \pL interaction.

One important aspect is to verify the compatibility of this result with the scattering data. This is done by using the 12 data points for \pL elastic cross section below \ks of 180~\MeVc, corresponding to $p_\mathrm{lab}\approx400~\MeVc$, published in~\cite{Alexander:pLambda,Sechi-Zorn:pLambda} and used by the chiral calculations~\cite{Haidenbauer:NLO13,Haidenbauer:NLO19}. Using the Usmani parameterizations within CATS, the corresponding total cross sections have been evaluated and compared to the scattering data. The Usmani potential tuned to NLO19 ($f_1=1.41~$fm) provides $\chi^2/$NDP=4.5/12, while the parameterization corresponding to the best femtosopic fit ($f_1=1.15~$fm) yields $\chi^2/$NDP=16.0/12, corresponding to 2.6$\sigma$ discrepancy. The evaluation of the cross section is based only under the consideration of the s-wave, thus a slight systematic underestimation of the total value is possible. Nevertheless, proceeding with the adopted procedure, if the Usmani repulsive core is modified to result in $f_1=1.24~$fm the compatibility to both scattering and femtoscopy data is 1.2~$\sigma$. This observation suggests that the two data sets can be used in a complementary way to constrain the theory. 

We would like to conclude this discussion with several important remarks related to the interpretation of the results on the \pL interaction presented in this work. The femtoscopic data and the adopted modeling of the source, using CECA, show a clear trend that the overall strength of the attraction should be lower compared to the currently accepted values. We have only modified the S=1 channel, as the correlation function is not particularly sensitive to the individual spin channels, thus the obtained result of $f_1=1.15$~fm serves as a lower limit within the triplet channel. We have anchored the Usmani potential to the NLO19 calculation, whereas the newer N$^2$LO calculation predicts a slightly reduced scattering length in the singlet channel~\cite{Haidenbauer:NNLO}. Moreover, the N$^2$LO has differences in the effective range, compared to NLO19, introducing further systematic bias to the exact value of $f_1$. In addition, the Usmani potential does not include the \pL coupling to \NS, with a kinematic threshold of $\ks\approx290~\MeVc$, and considers only s-waves. This is not an issue for the present analysis, as the upper fit range is purposefully selected to be only 180~\MeVc, nevertheless a small systematic bias may still be present. For these reasons, the best future strategy is to use both of the scattering and femtoscopic data to constrain the chiral theory directly, without using the Usmani potential as a proxy. Given the new theoretical developments, in particular the N$^2$LO calculation~\cite{Haidenbauer:NNLO}, it is indeed very interesting to perform such a dedicated study, utilizing the CECA framework.

%% file: Chapters/Summary.tex
\section{Summary and outlook}\label{sec:Summary}

A numerical framework called CECA has been developed and presented in this work. It can effectively model particle emission in small collision systems. The framework is based on single-particle emission, where the particles are treated as point-like objects with well-defined space and momentum components. The properties of hadronization are described by three effective parameters, which take into account random fluctuations, geometrical space-momentum correlations representing a collective expansion of the collision system, and a time evolution parameter. Additionally, CECA requires the input of the individual momentum distributions for each simulated species, as well as the amount and type of short-lived resonances that feed into the particles of interest.

The new framework has been used to analyze the \mt-differentially measured \pP and \pL correlation functions from the ALICE collaboration. It has been shown that a common emission source for both particles is consistent with the data, but only given a \pL scattering length within the triplet channel of $f_1=1.15\pm0.07~$fm. The existing scattering data is in slight tension with this value, nevertheless, $f_1=1.24~$fm is equally compatible with the femtoscopic and scattering data, deviating from both by only 1.2$\sigma$. These $f_1$ values are significantly lower compared to the chiral effective field theory, which is evaluated at NLO and N$^2$LO and is currently tuned only to the existing scattering data. Undoubtedly, the best future strategy to constrain theoretical models is the combined usage of femtoscopic and scattering data, which seem to provide complementary constraints.

The study of the emission source greatly benefits from correlations between identical pions, due to their quantum statistical properties and lack of significant strong final state interaction. Nevertheless, their description in small collision systems is complicated, due to the large kinematic effects of feed-down from resonances. If the cocktail of relevant resonances is well constrained, along with the corresponding momenta distributions, the CECA framework can be a vital tool to describe existing and future pion analyses in small collision systems realized at high energies. Such studies will complement the baryon--baryon analysis, and provide a more clear picture of the assumption of a common emission source. 
Moreover, the effective parameterization used in CECA can help to constrain advanced transport models and eliminate ambiguities in the interpretation of correlation results. 

The CECA framework is not limited to modeling only the two-body source function and can be used for the description of many-body sources. This feature will become particularly useful during the third data-taking period of the LHC, as there are existing plans to employ femtoscopy to investigate the genuine three-body force~\cite{ALICE:3B}. This requires knowledge of the two- and three-body source functions, both of which can be consistently modeled within CECA.

%% file: Chapters/Appendix.tex
\section*{Appendix}\label{app}

\begin{figure*}[ht]
    \centering
    \includegraphics[width=0.32\textwidth]{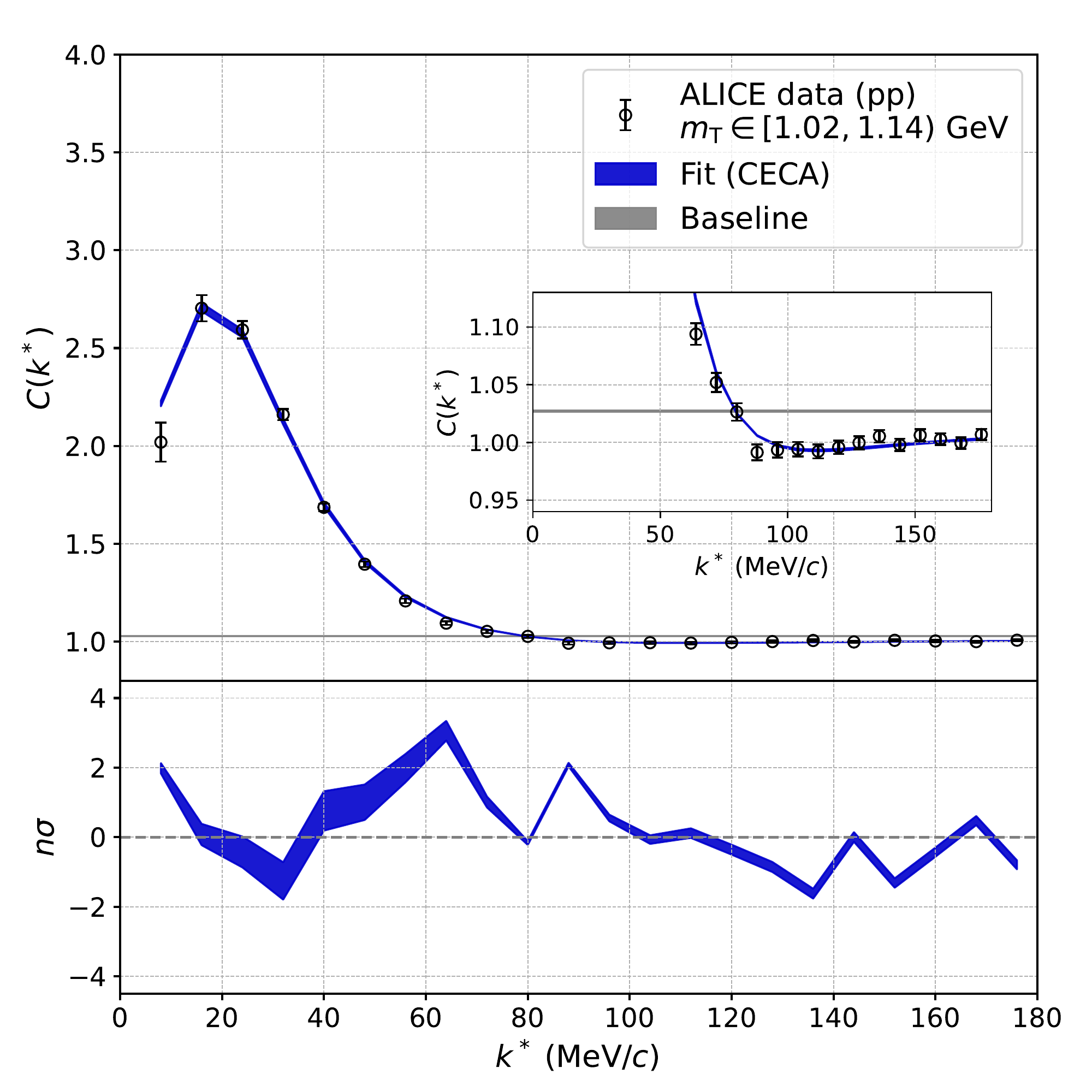}
    \includegraphics[width=0.32\textwidth]{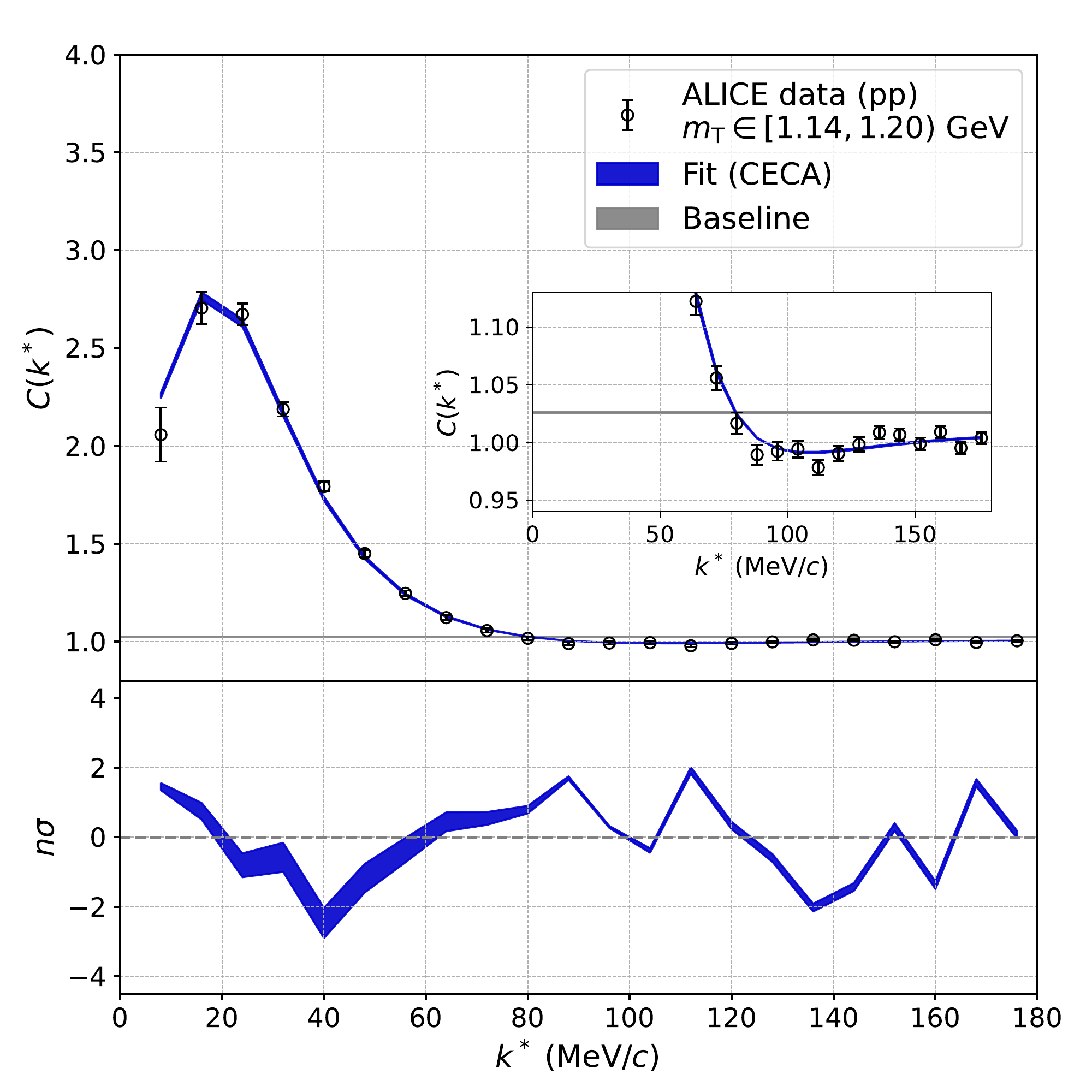}
    \includegraphics[width=0.32\textwidth]{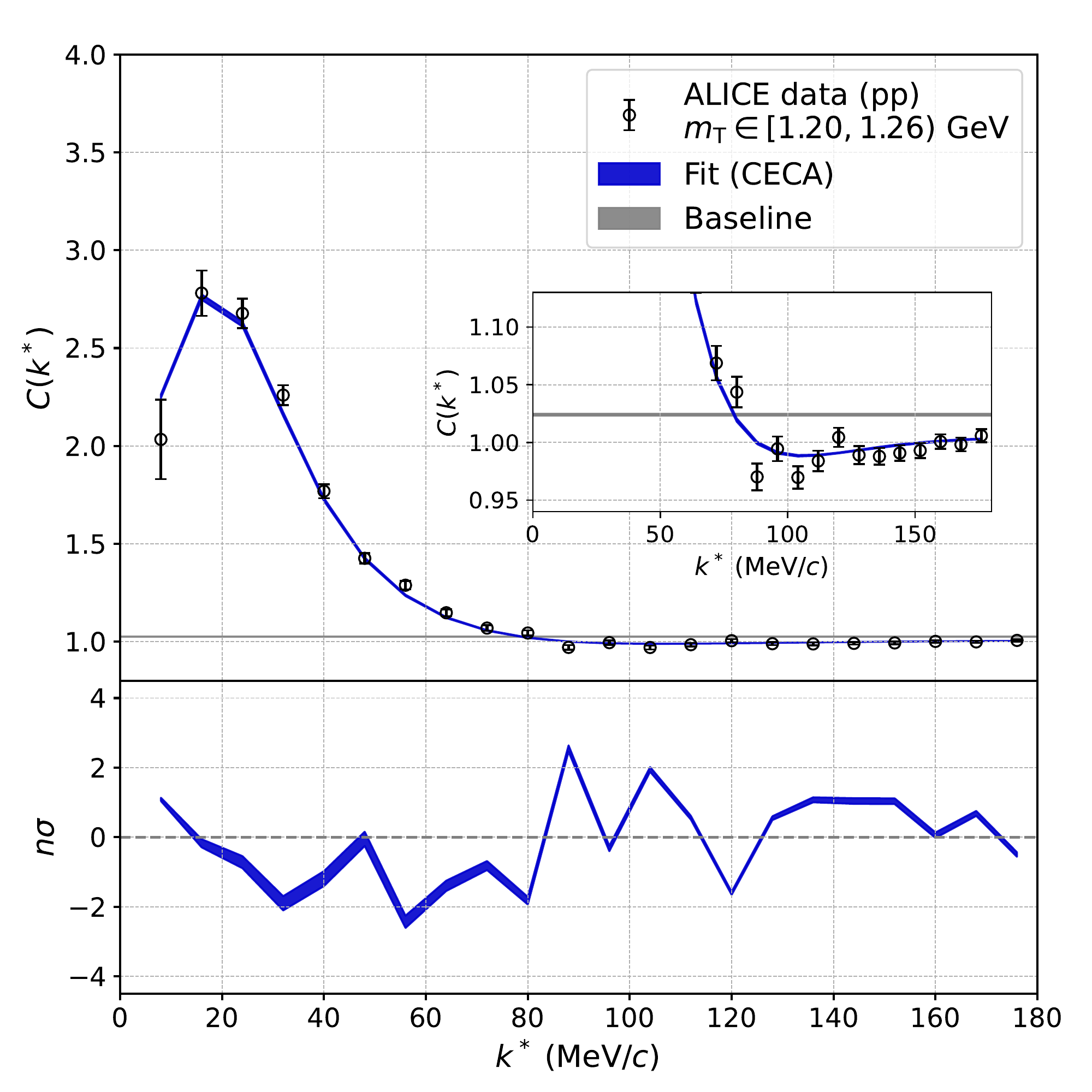}\\
    \includegraphics[width=0.32\textwidth]{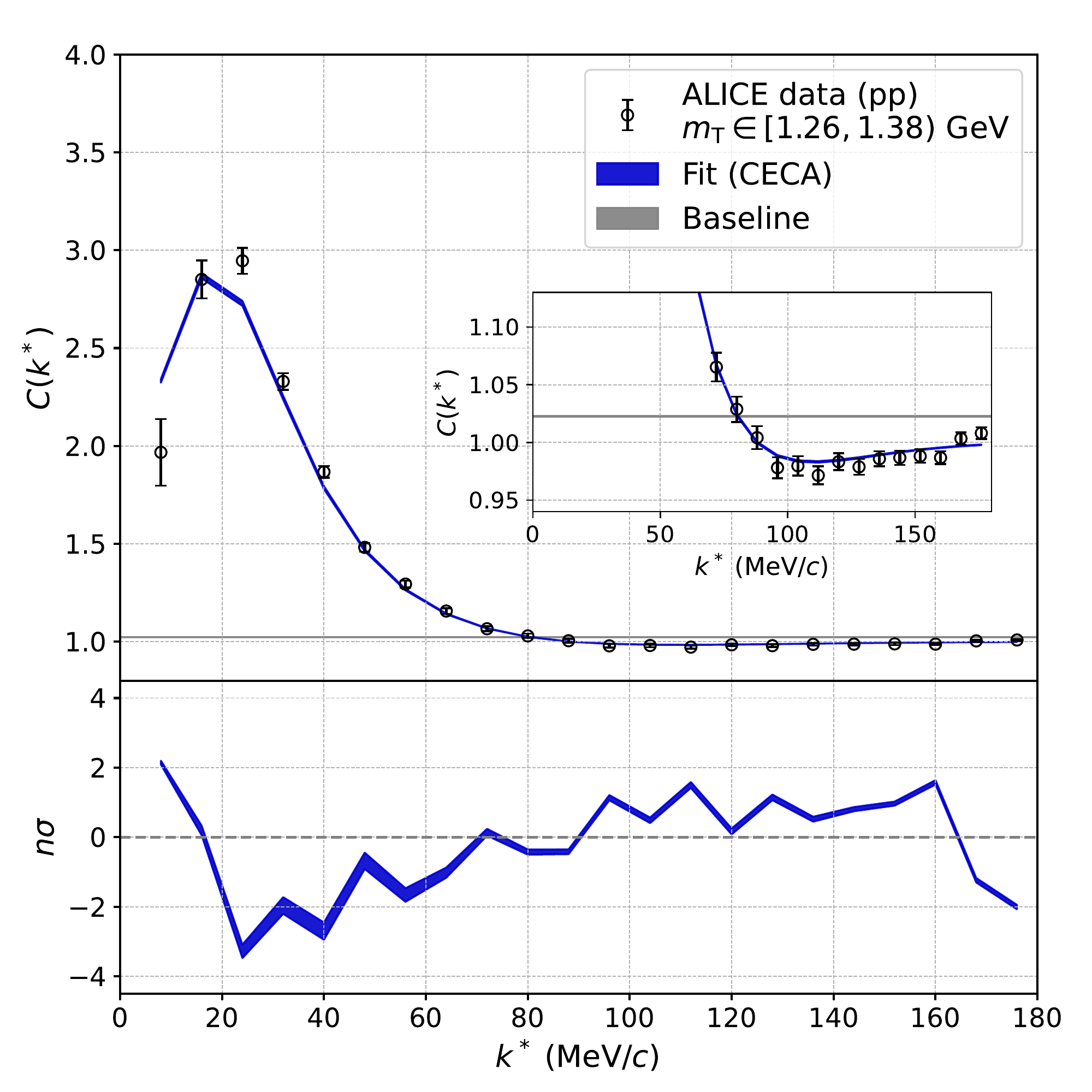}
    \includegraphics[width=0.32\textwidth]{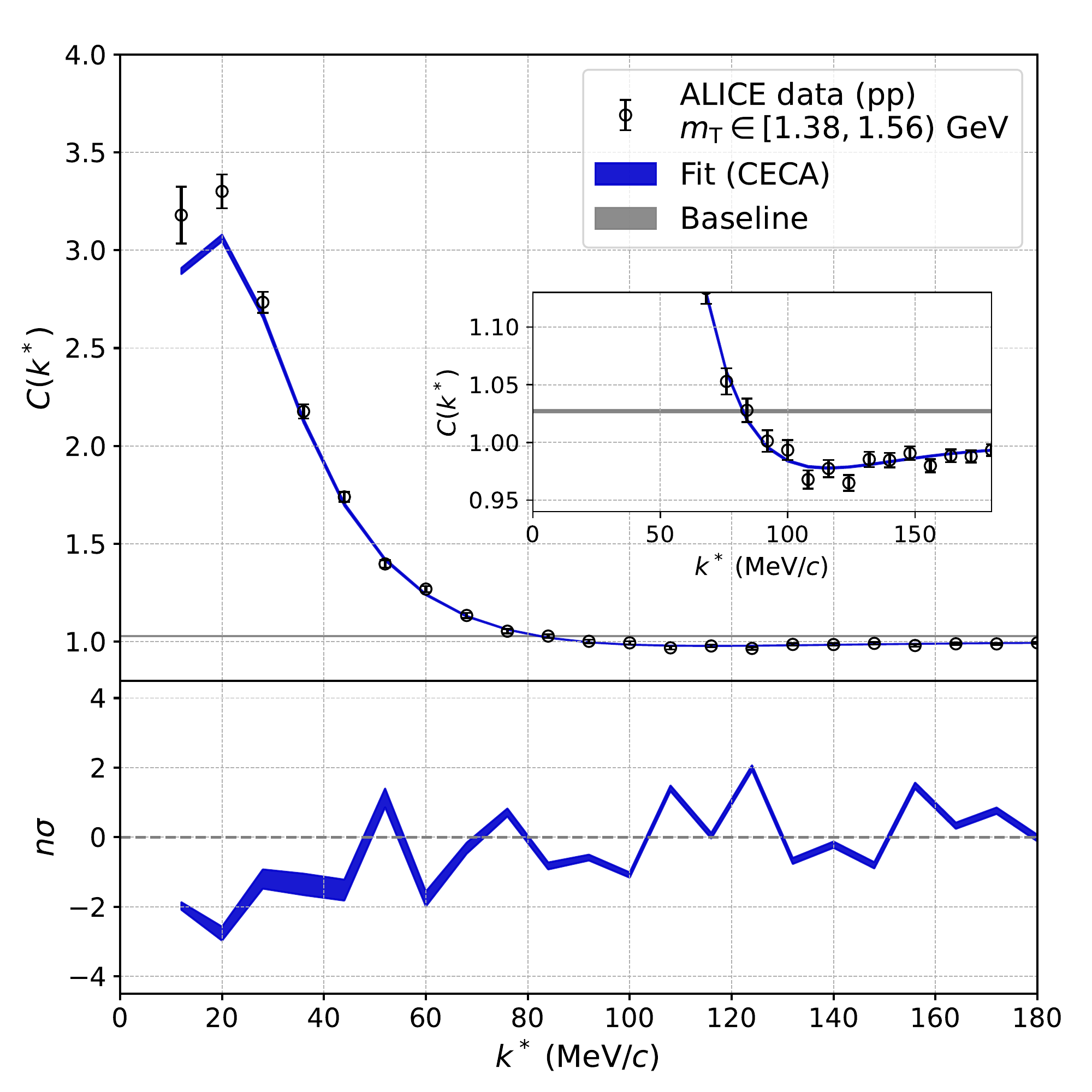}\\
    \includegraphics[width=0.32\textwidth]{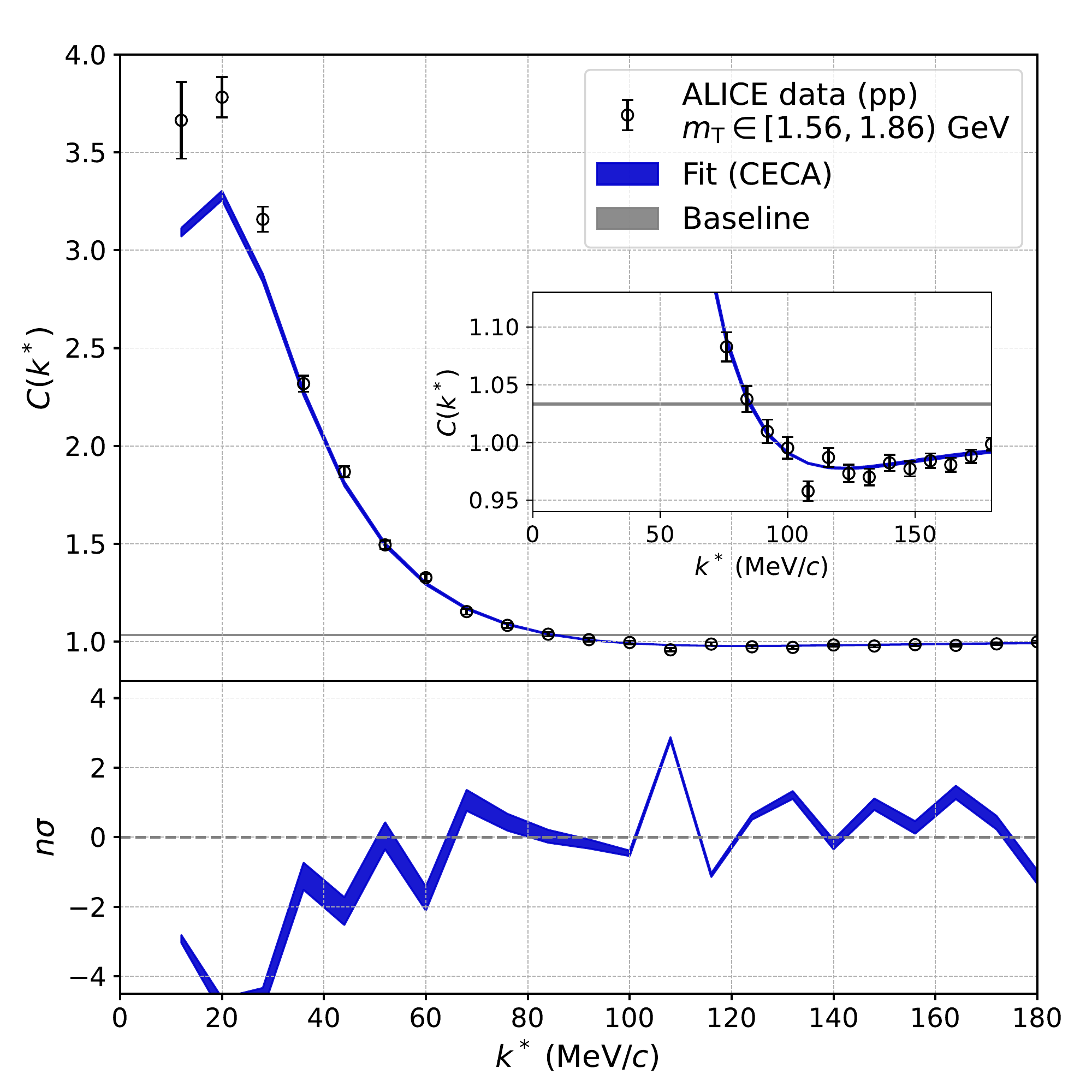}
    \includegraphics[width=0.32\textwidth]{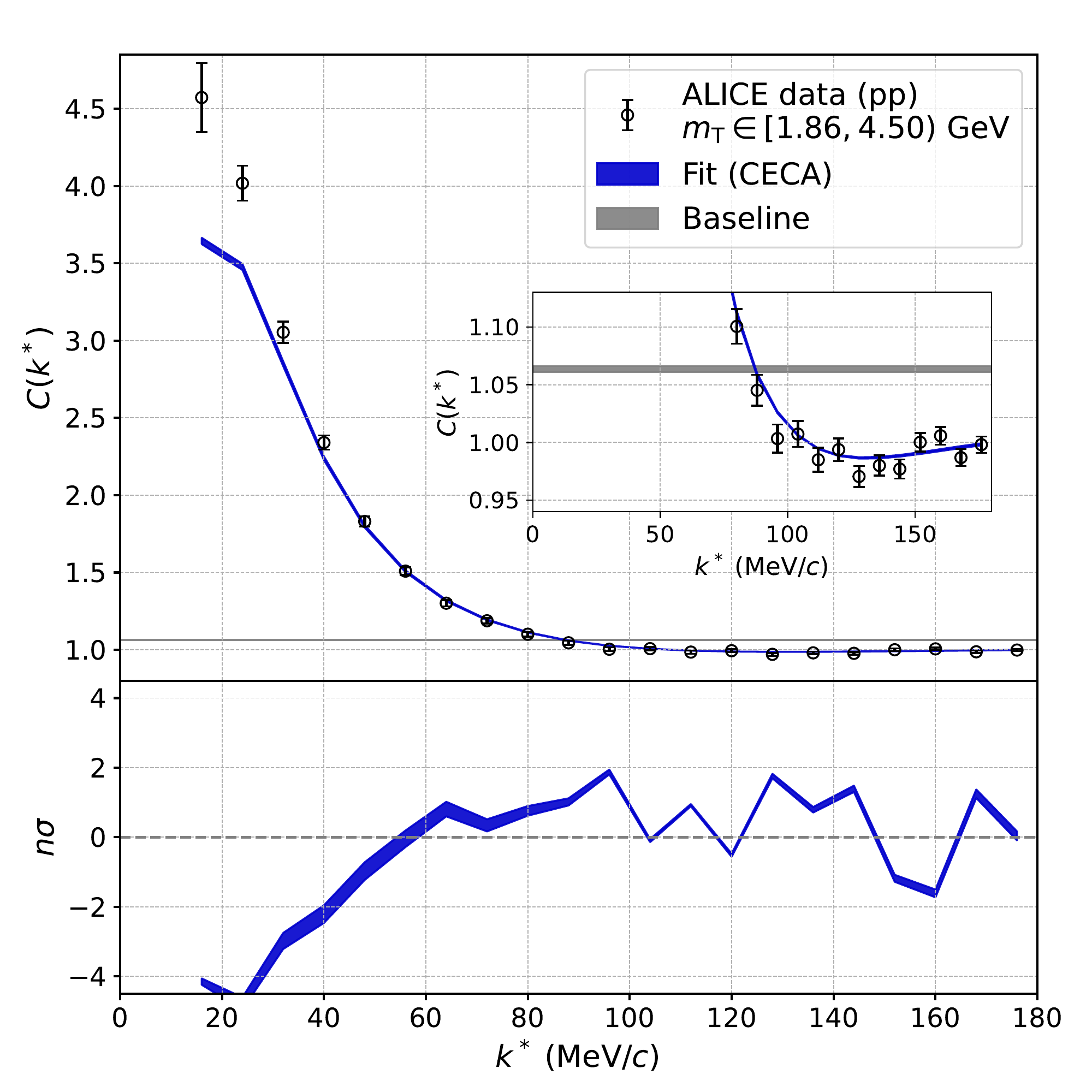}
    \caption[]{The \pP correlation functions (in all \mt bins) fitted using the CECA source. The \pL interaction, which is relevant for the feed-down into \pP, is taken from the fit results shown in Fig.~\ref{fig:ana:Ck_pL_Usm_ALL}.}
    \label{fig:ana:Ck_pp_Usm_ALL}
\end{figure*}

\begin{figure*}[ht]
    \centering
    \includegraphics[width=0.32\textwidth]{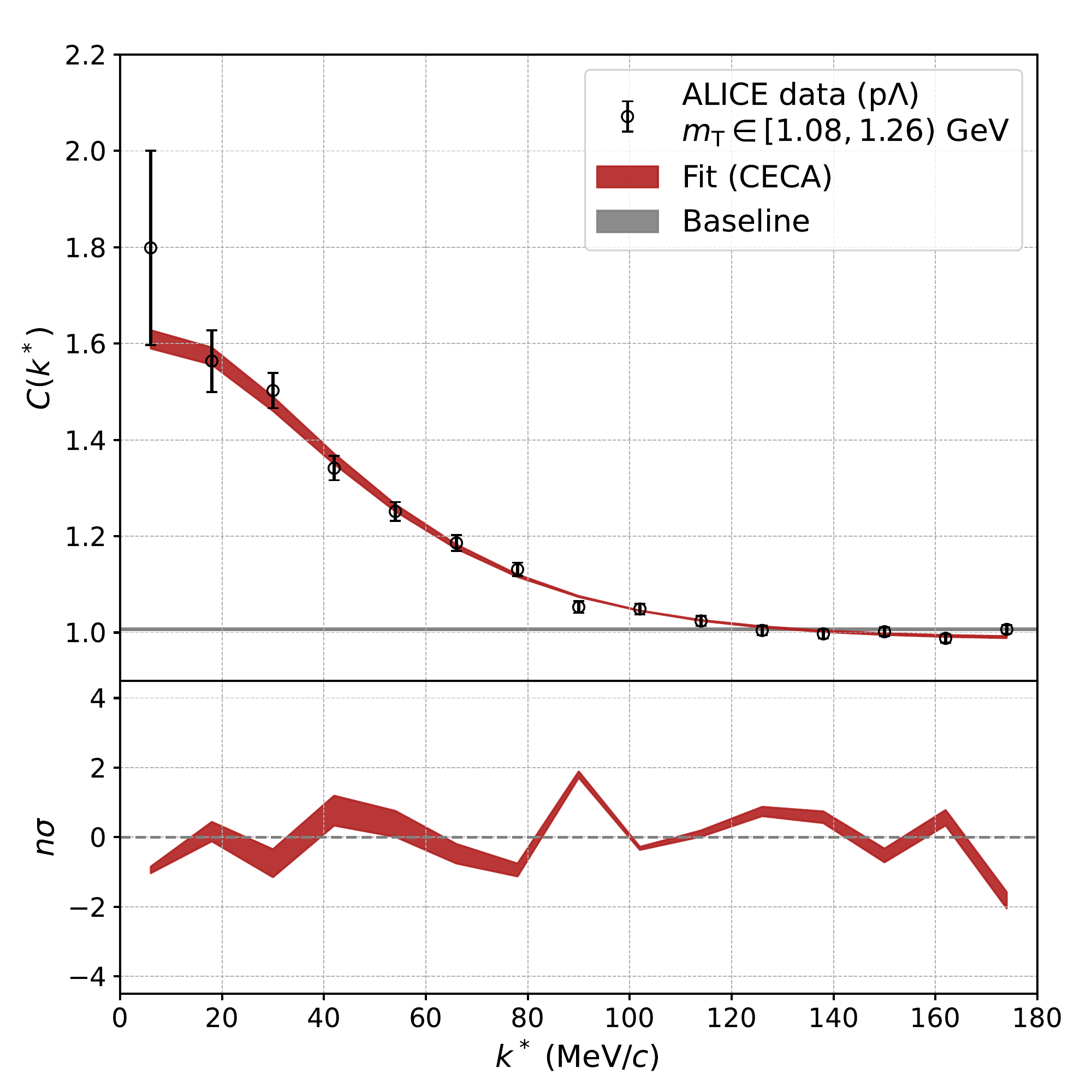}
    \includegraphics[width=0.32\textwidth]{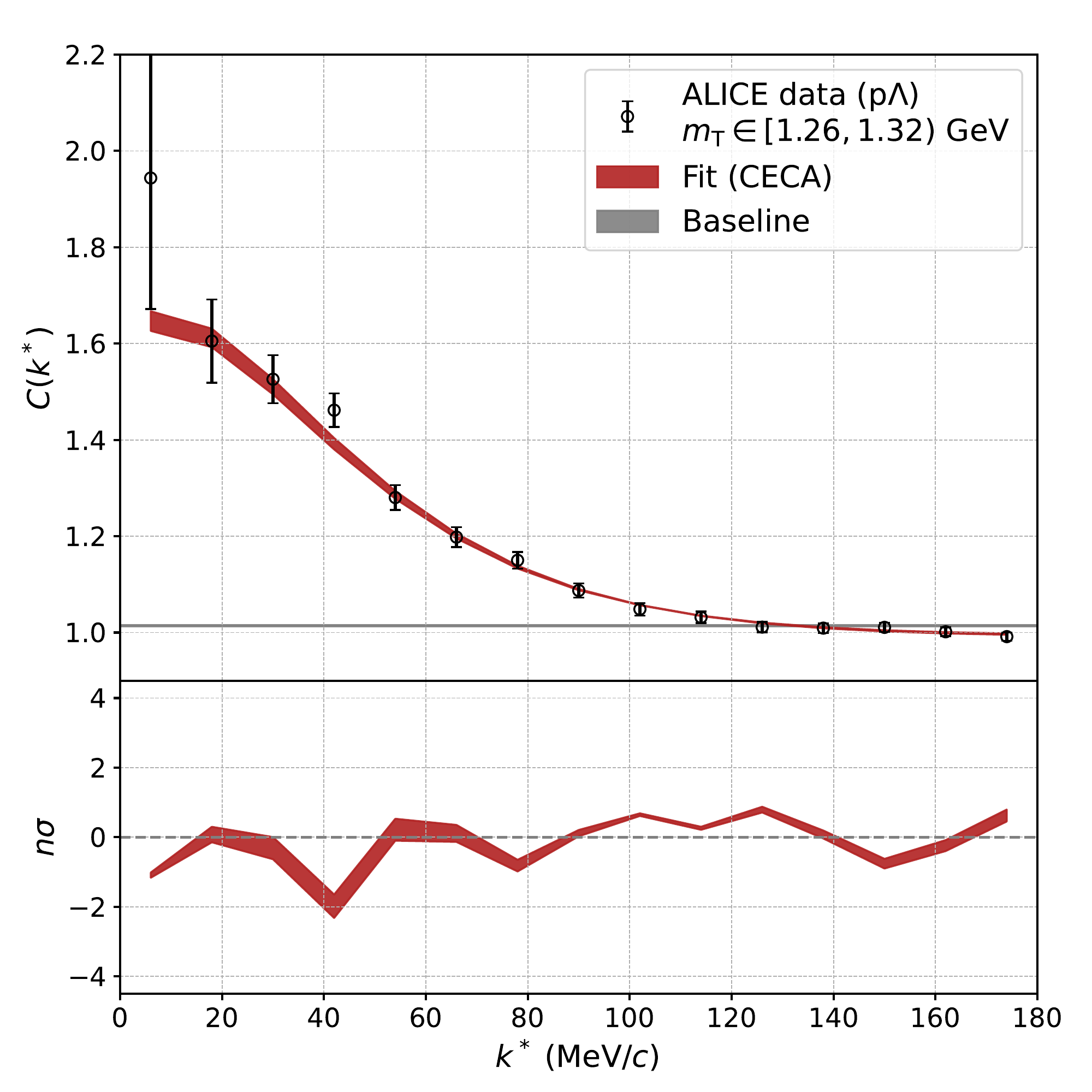}
    \includegraphics[width=0.32\textwidth]{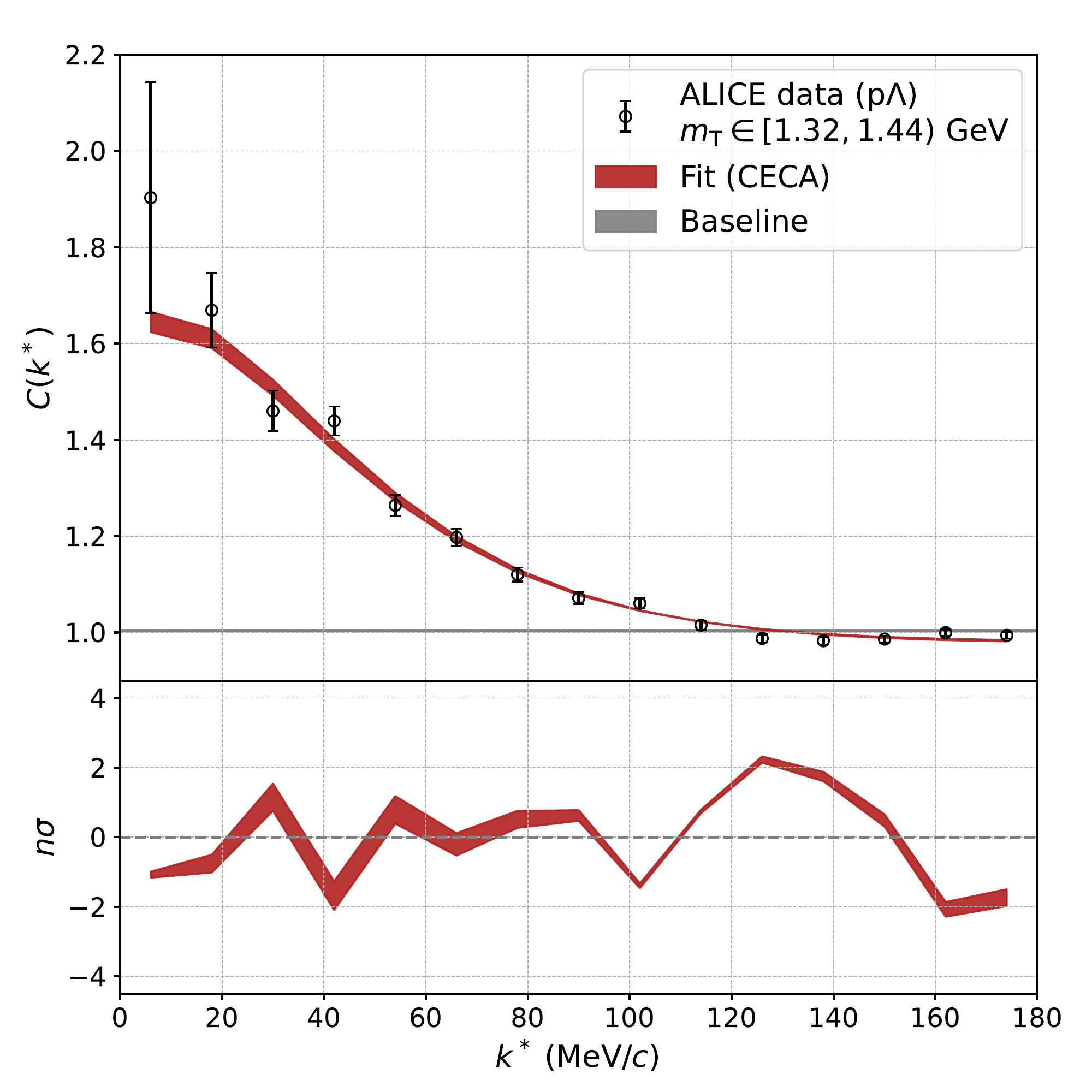}\\
    \includegraphics[width=0.32\textwidth]{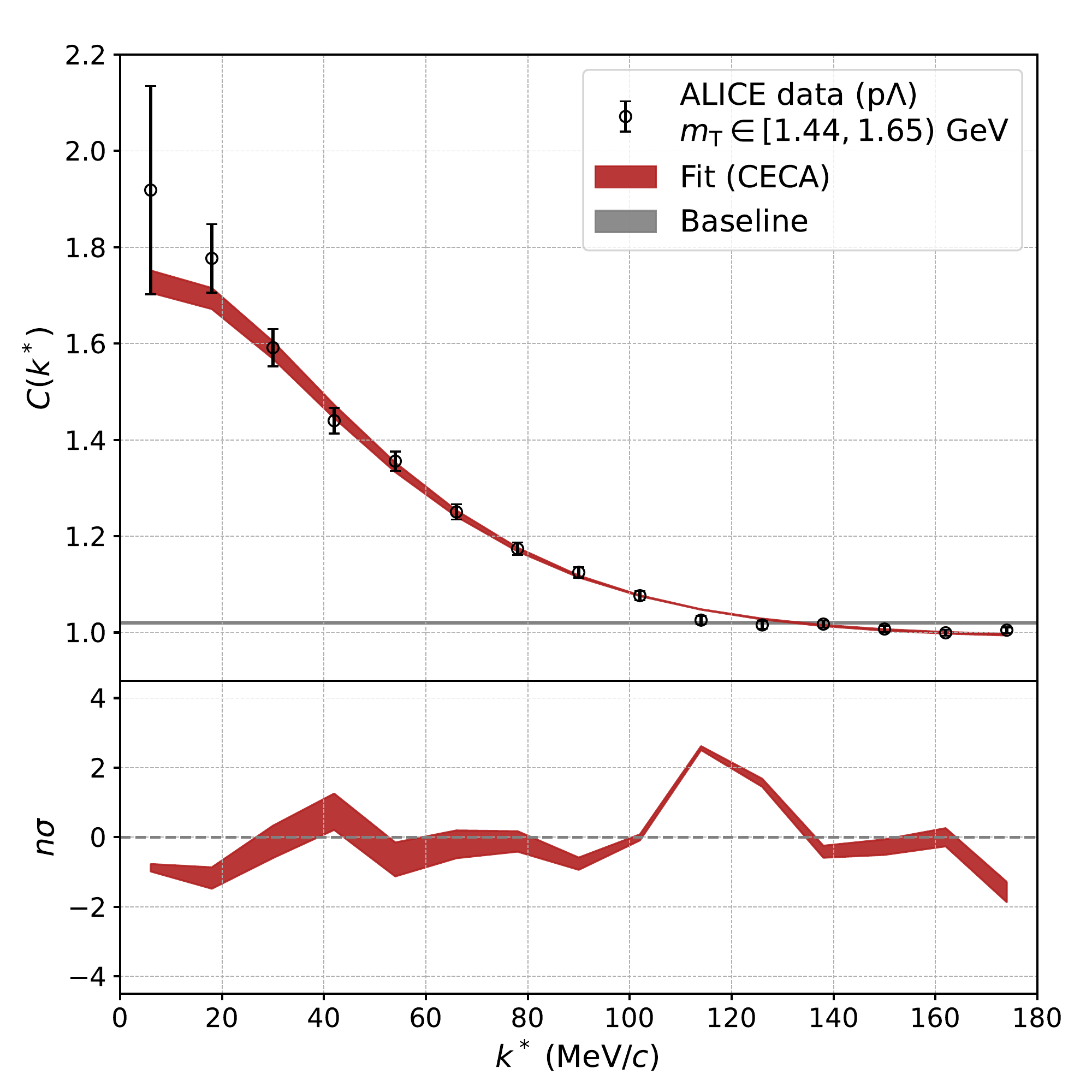}
    \includegraphics[width=0.32\textwidth]{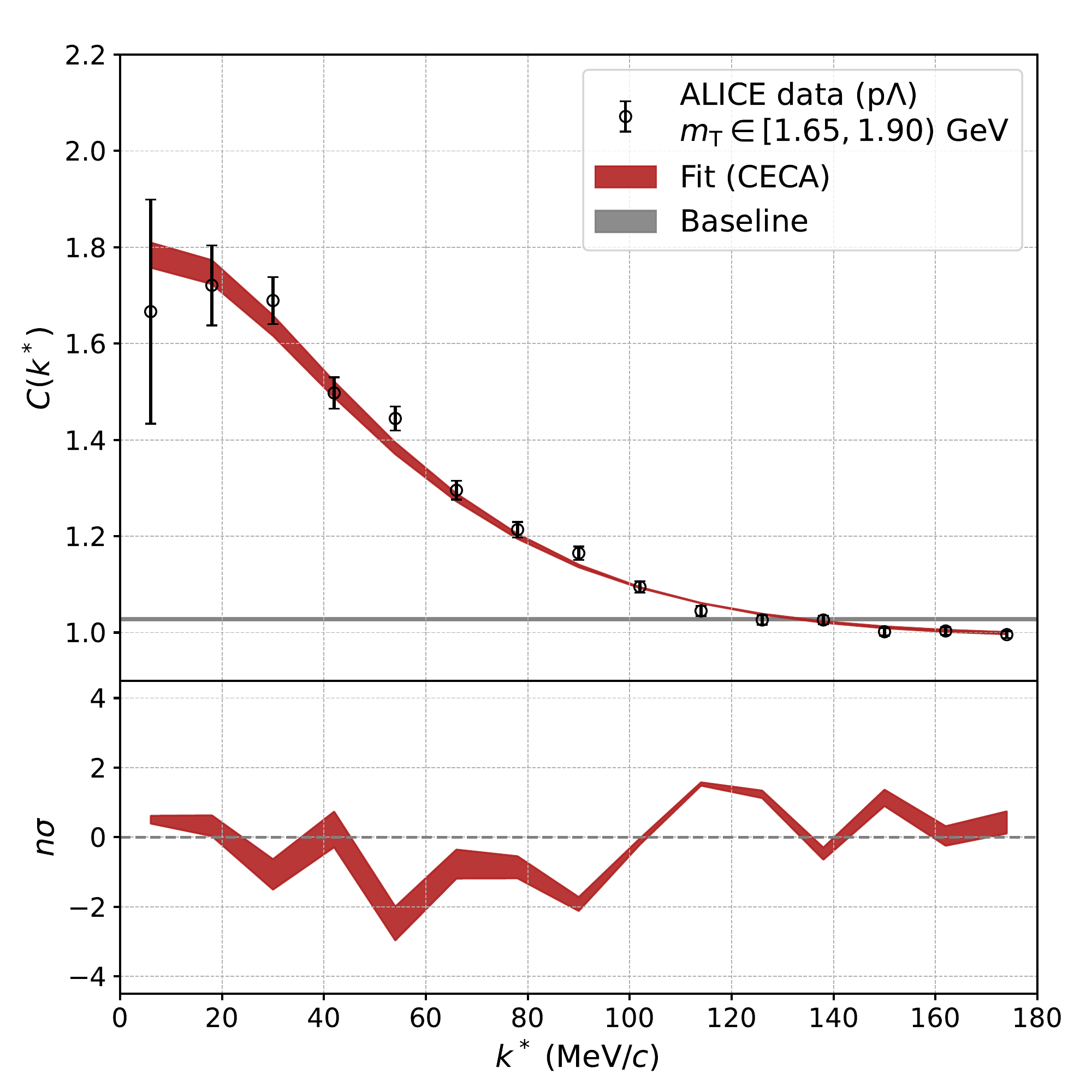}
    \includegraphics[width=0.32\textwidth]{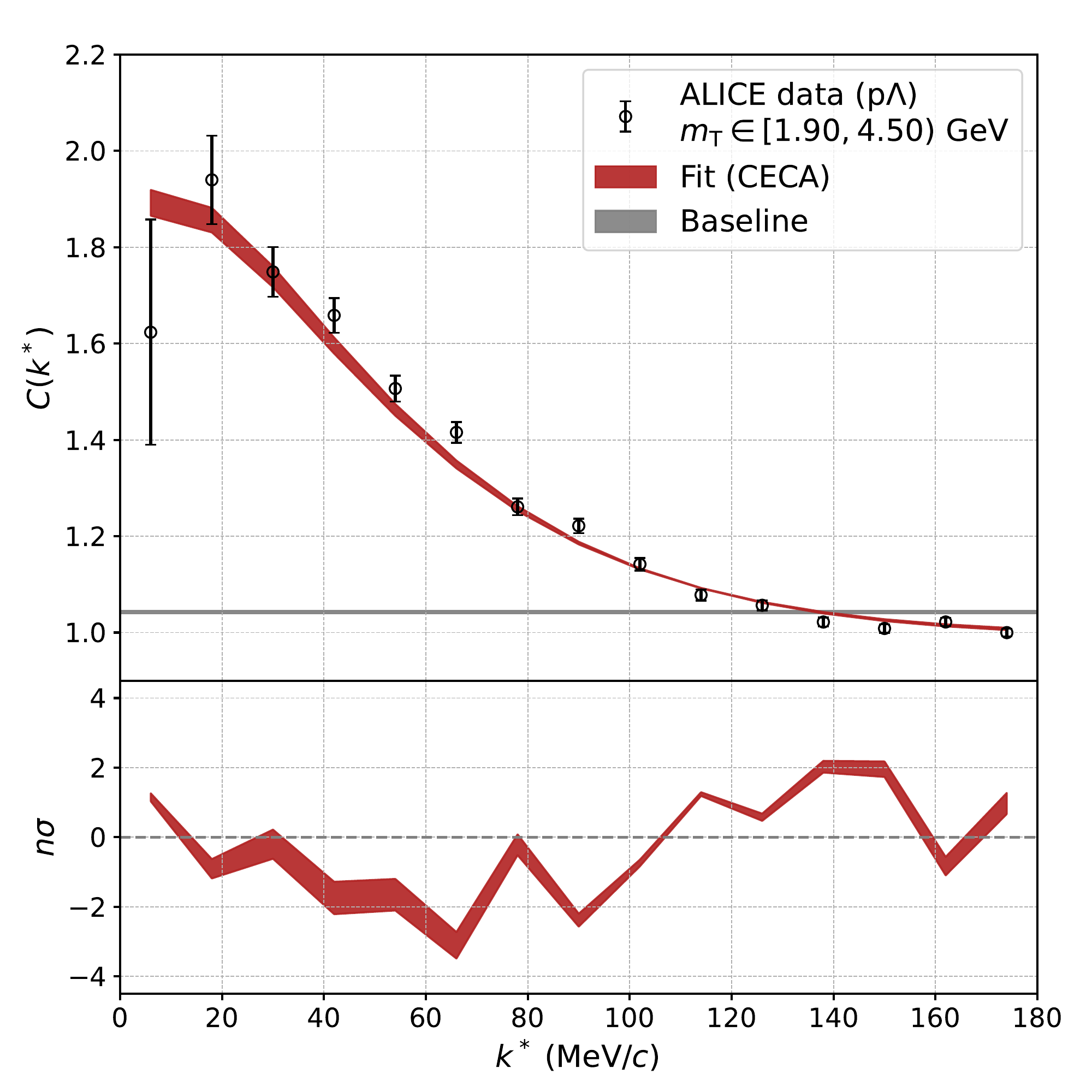}
    \caption[]{The \pL correlation functions (in all \mt bins) fitted using the CECA source, where the interaction is modeled using the Usmani potential. The repulsive core parameters are left free in the fit.}
    \label{fig:ana:Ck_pL_Usm_ALL}
\end{figure*}

\begin{figure*}[ht]
    \centering
    \includegraphics[width=0.32\textwidth]{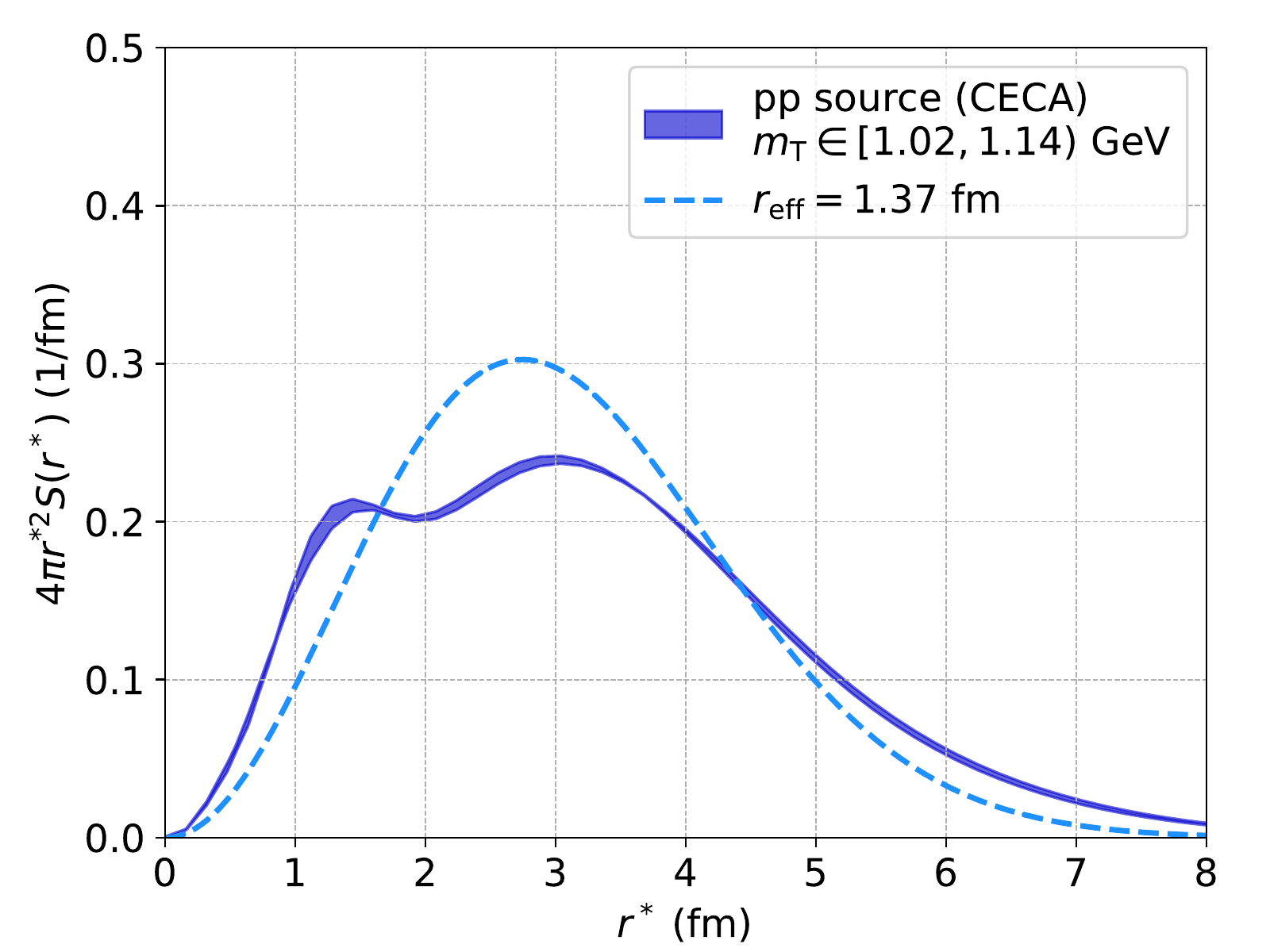}
    \includegraphics[width=0.32\textwidth]{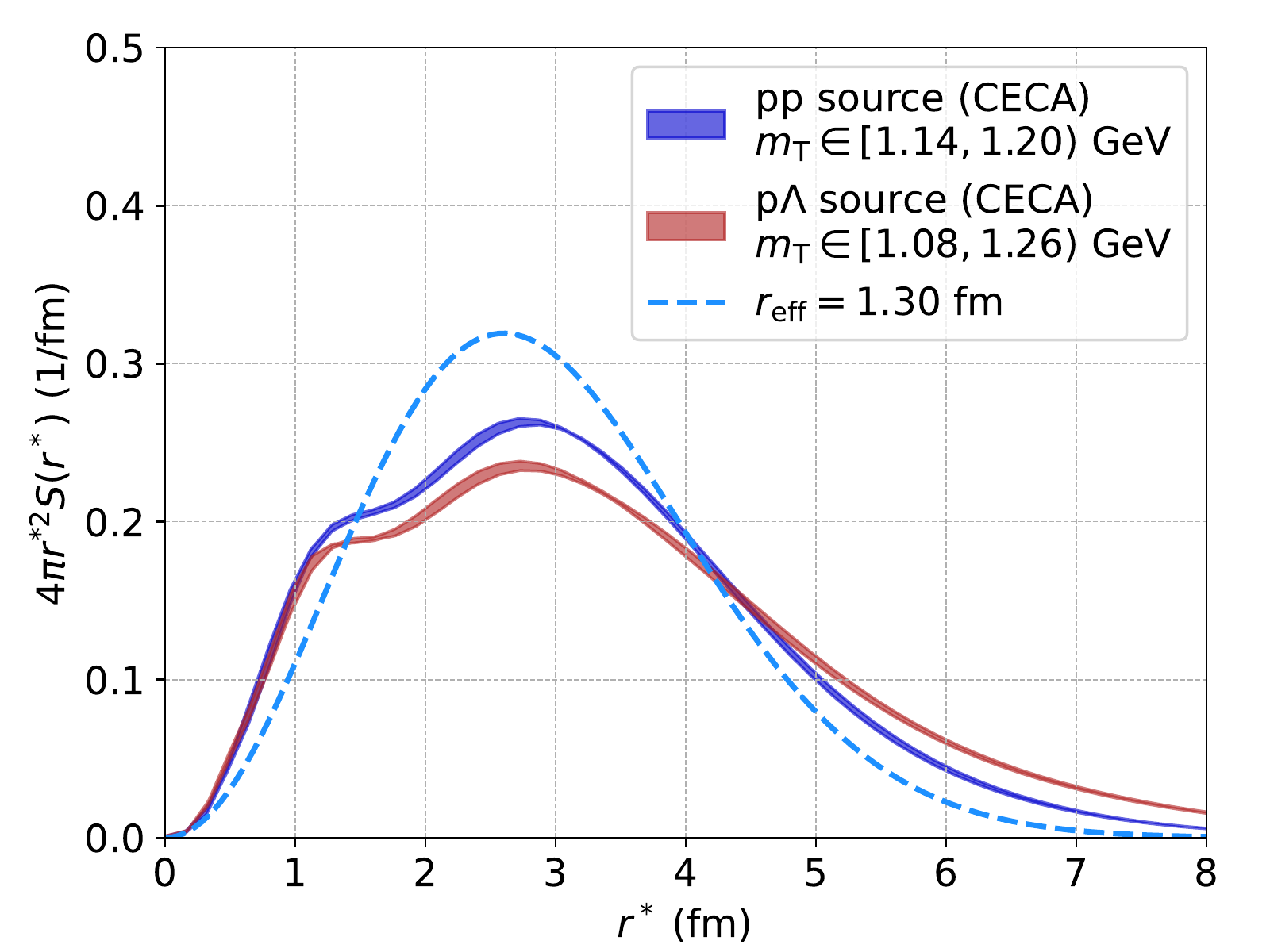}
    \includegraphics[width=0.32\textwidth]{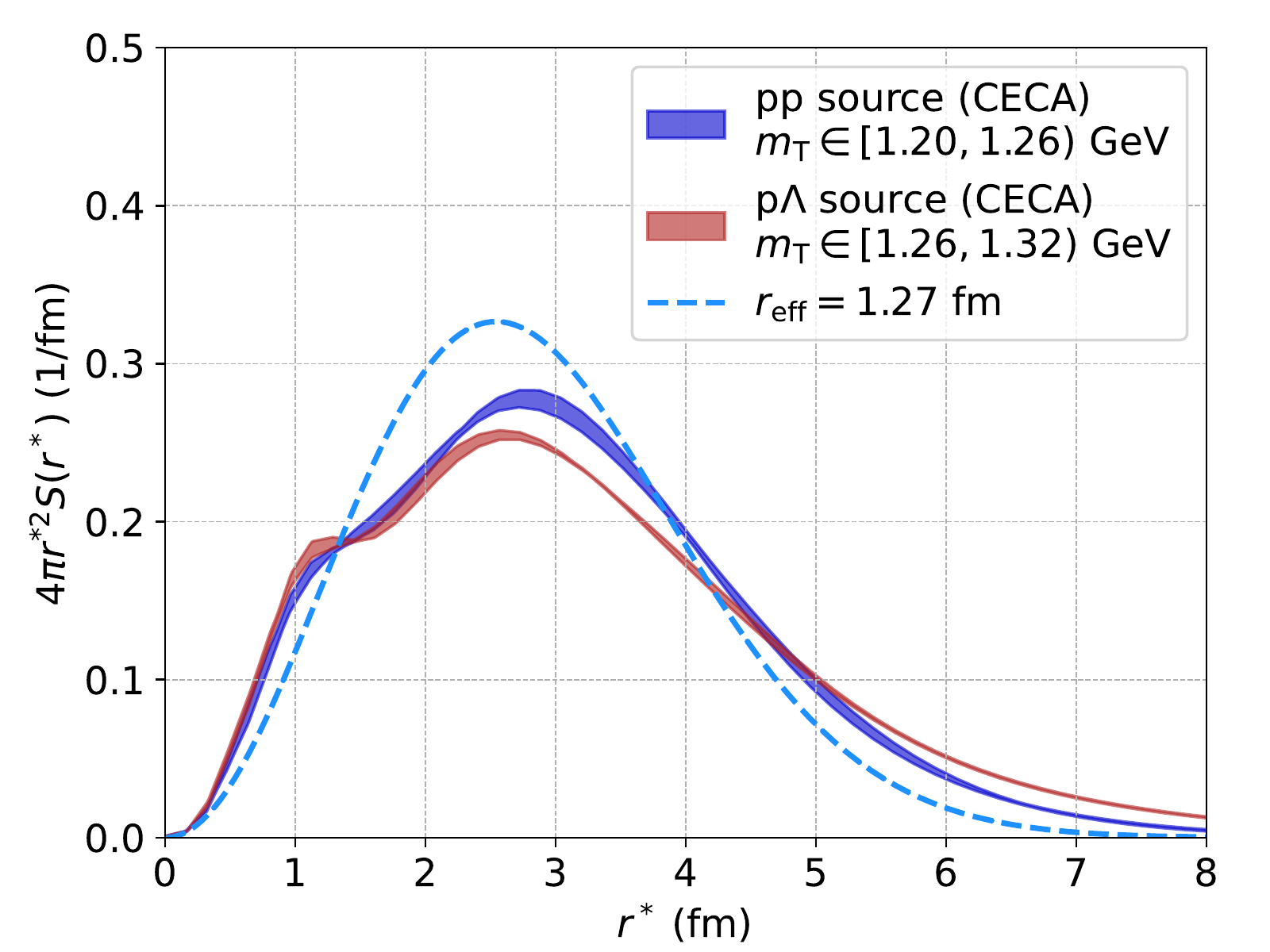}\\
    \includegraphics[width=0.32\textwidth]{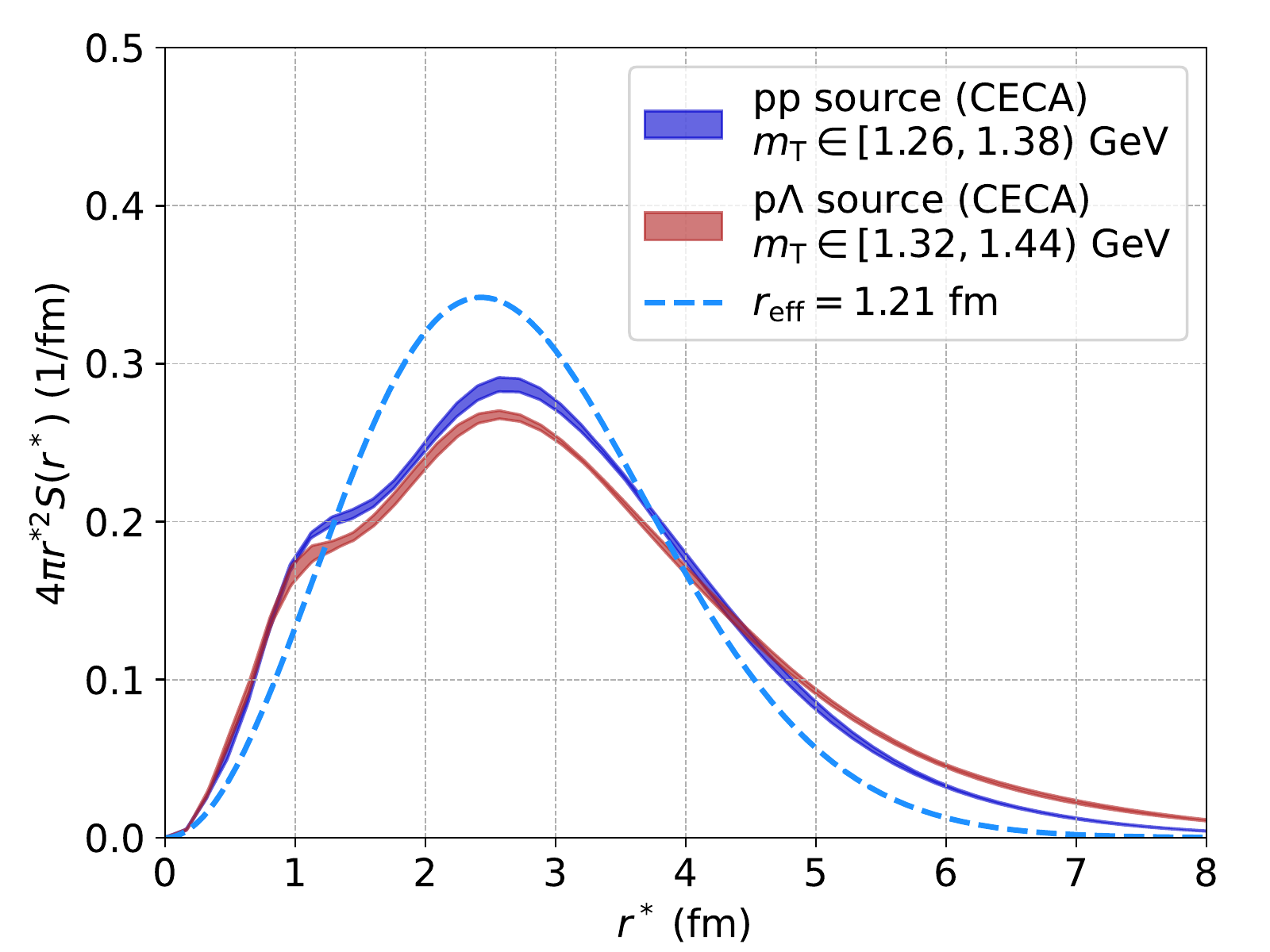}
    \includegraphics[width=0.32\textwidth]{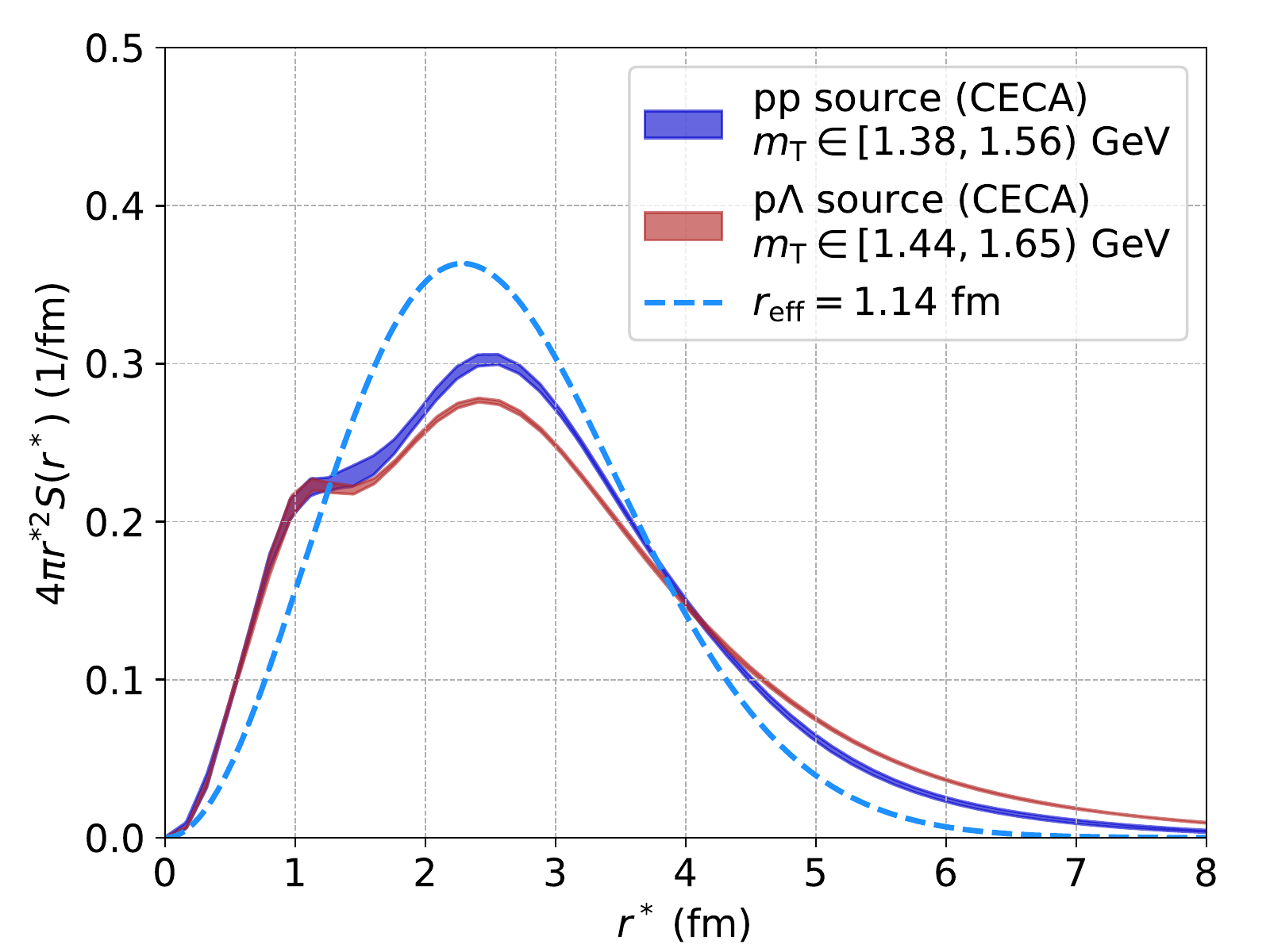}\\
    \includegraphics[width=0.32\textwidth]{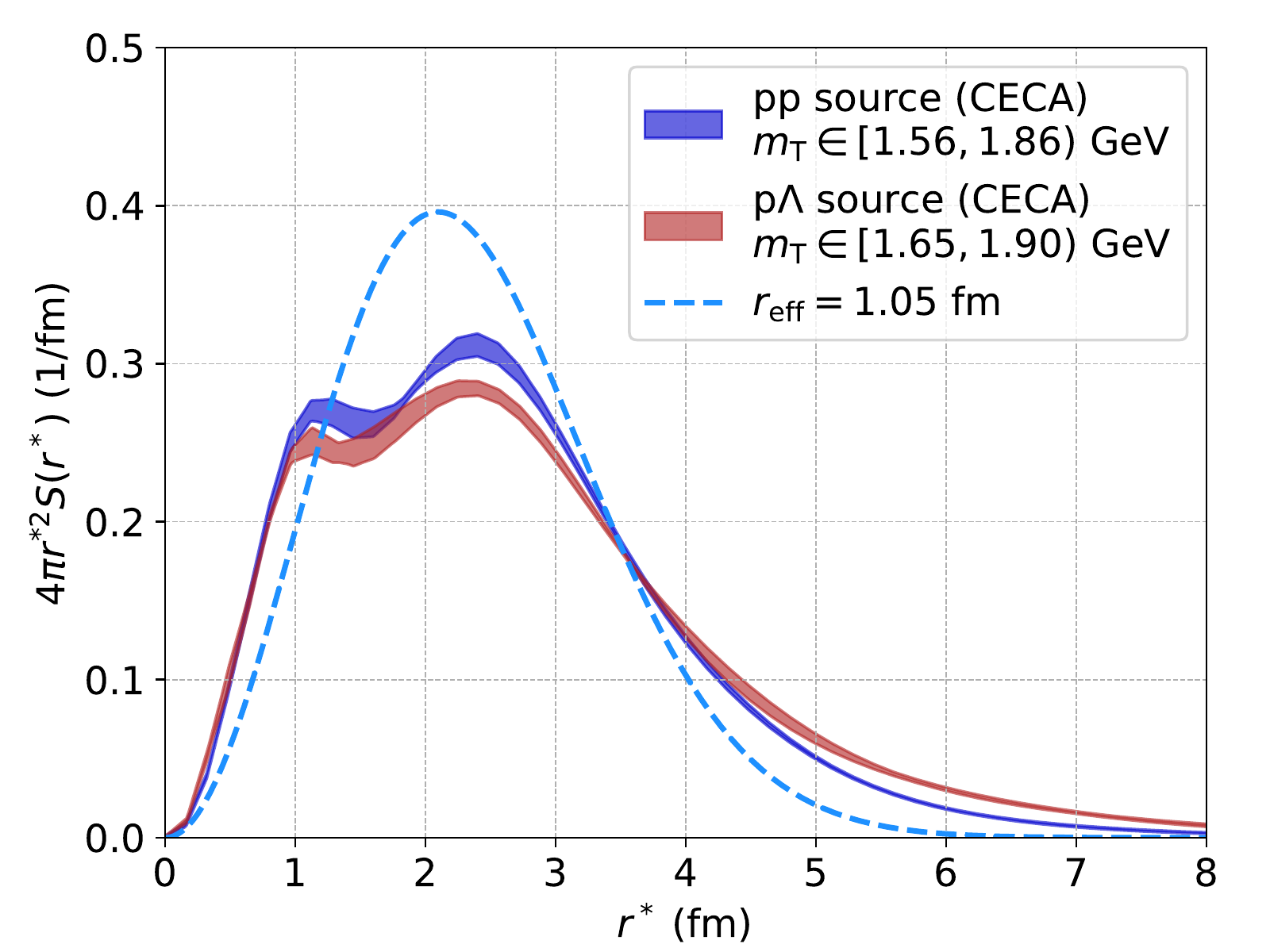}
    \includegraphics[width=0.32\textwidth]{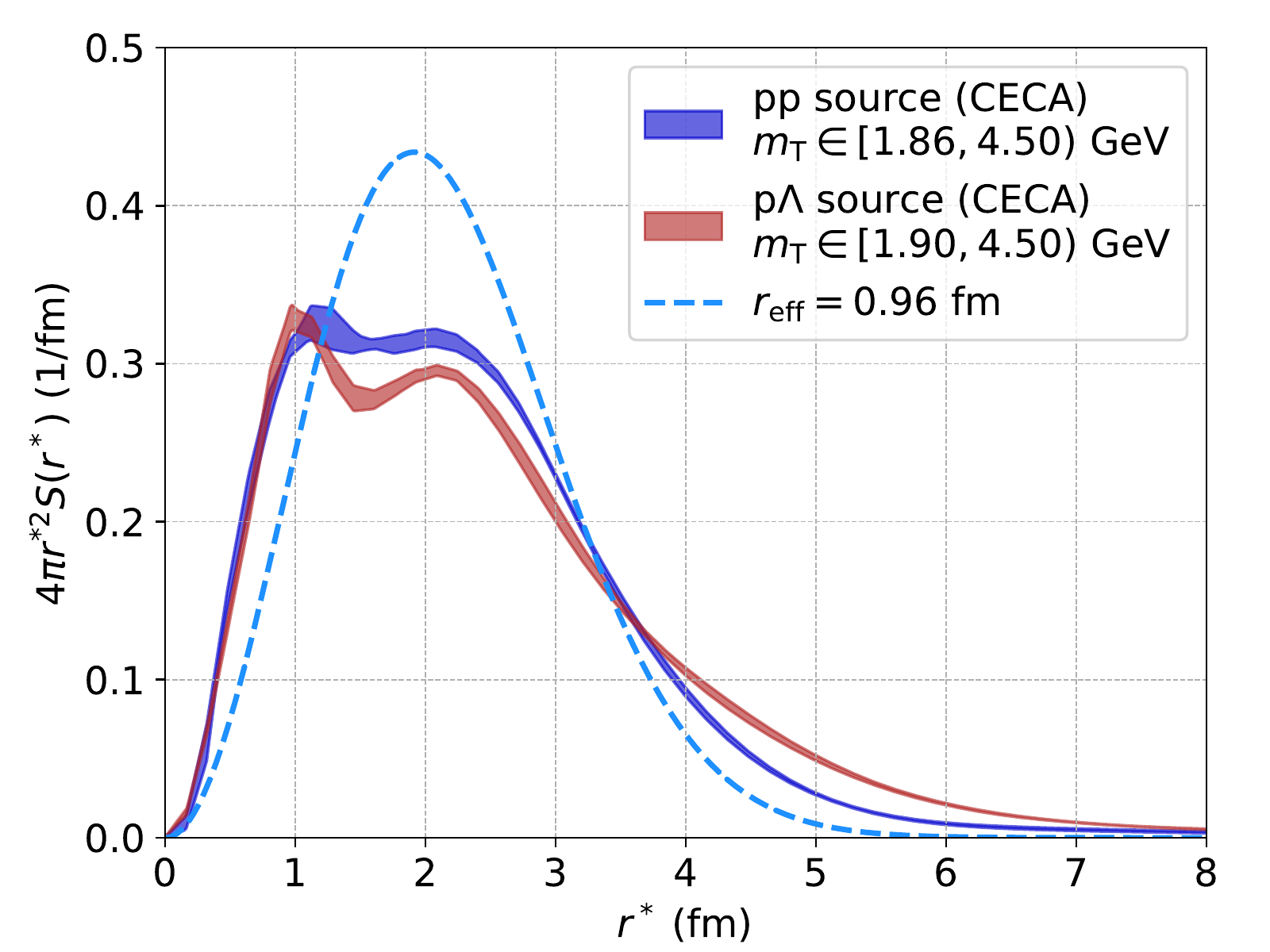}
    \caption[]{The \pP (blue) and \pL (dark red) total source functions in each \mt bin. The \pL interaction is modeled by the Usmani potential and the relevant parameters of the repulsive core fitted to the data. The dashed line corresponds to the effective \pP Gaussian source ($r_\mathrm{eff}$) extracted by the ALICE collaboration, by analyzing the same data~\cite{ALICE:Source}.}
    \label{fig:ana:Sr_pp_Usm_ALL}
\end{figure*}

Figure~\ref{fig:ana:Ck_pp_Usm_ALL} (\ref{fig:ana:Ck_pL_Usm_ALL}) shows the fits to the \pP (\pL) correlation functions, using the Usmani potential with free parameters of the repulsive core.  The corresponding CECA source distributions are plotted in Fig.~\ref{fig:ana:Sr_pp_Usm_ALL}.

